\UseRawInputEncoding
\documentclass[sigconf]{acmart}

\usepackage{enumitem}
\usepackage{subfigure}
\usepackage{multirow}

\AtBeginDocument{%
  \providecommand\BibTeX{{%
    \normalfont B\kern-0.5em{\scshape i\kern-0.25em b}\kern-0.8em\TeX}}}

\copyrightyear{2021}
\acmYear{2021}
\setcopyright{acmcopyright}\acmConference[KDD '21]{Proceedings of the 27th ACM SIGKDD Conference on Knowledge Discovery and Data Mining}{August 14--18, 2021}{Virtual Event, Singapore}
\acmBooktitle{Proceedings of the 27th ACM SIGKDD Conference on Knowledge Discovery and Data Mining (KDD '21), August 14--18, 2021, Virtual Event, Singapore}
\acmPrice{15.00}
\acmDOI{10.1145/3447548.3467077}
\acmISBN{978-1-4503-8332-5/21/08}

\begin{document}

\title{An Embedding Learning Framework for Numerical Features in CTR Prediction}

\author{Huifeng Guo$^{1*}$, Bo Chen$^{1*}$, Ruiming Tang$^{1}$, Weinan Zhang$^{2}$, Zhenguo Li$^{1}$, Xiuqiang He$^{1}$}
\thanks{*Co-first authors with equal contributions. Huifeng Guo and Ruiming Tang are the corresponding authors.}
\email{{huifeng.guo, chenbo116, tangruiming, li.zhenguo, hexiuqiang1}@huawei.com; wnzhang@sjtu.edu.cn}

\affiliation{%
  \institution{
  $^{1}$Huawei Noah's Ark Lab~~~$^{2}$Shanghai Jiao Tong University
  }
  \country{}
}

\begin{abstract}
Click-Through Rate (CTR) prediction is critical for industrial recommender systems, where most deep CTR models follow an Embedding \& Feature Interaction paradigm. However, the majority of methods focus on designing network architectures to better capture feature interactions while the feature embedding, especially for numerical features, has been overlooked. Existing approaches for numerical features are difficult to capture informative knowledge because of the low capacity or hard discretization based on the offline expertise feature engineering.
In this paper, we propose a novel embedding learning framework for numerical features in CTR prediction (AutoDis) with \textit{high model capacity}, \textit{end-to-end training} and \textit{unique representation} properties preserved. AutoDis consists of three core components: \emph{meta-embeddings}, \emph{automatic discretization} and \emph{aggregation}. Specifically, we propose meta-embeddings for each numerical field to learn global knowledge from the perspective of field with a manageable number of parameters. Then the differentiable automatic discretization performs soft discretization and captures the correlations between the numerical features and meta-embeddings. Finally, distinctive and informative embeddings are learned via an aggregation function.
Comprehensive experiments on two public and one industrial datasets are conducted to validate the effectiveness of AutoDis. Moreover, AutoDis has been deployed onto a mainstream advertising platform, where online A/B test demonstrates the improvement over the base model by 2.1\% and 2.7\% in terms of CTR and eCPM, respectively. In addition, the code of our framework is publicly available in MindSpore\footnote{\url{https://gitee.com/mindspore/mindspore/tree/master/model_zoo/research/recommend/autodis}}.



\end{abstract}

\ccsdesc[500]{Information systems~Information retrieval}
\ccsdesc[500]{Information systems~Recommender systems}
\ccsdesc[500]{Computing methodologies~Machine learning}

\keywords{Numerical Features, Embedding Learning, Click-Through Rate Prediction, Neural Network}


\maketitle

\section{Introduction}\label{sec:intro}

To alleviate the problem of information explosion, recommender systems are widely deployed to provide personalized information filtering in online information services, such as web search, commodity recommendation and online advertising. CTR prediction is crucial in recommender systems, which is to estimate the probability that a user will click on a recommended item under a specific context, so that recommendation decisions can be made based on the predicted CTR values~\cite{ftrl,din}.
Because of the superior performance of representation learning in computer vision and natural language processing, deep learning techniques attract the attention of recommendation community. Therefore, plenty of industrial companies propose various deep CTR models and deploy them in their commercial systems, such as Wide \& Deep~\cite{wide-deep} in Google Play, DeepFM~\cite{deepfm} in Huawei AppGallery and DIN~\cite{din} in Taobao.


\begin{figure*}[!t]
    \centering
    \setlength{\belowcaptionskip}{-0.4cm}
 	\setlength{\abovecaptionskip}{-0cm}
    \includegraphics[width=0.75\textwidth]{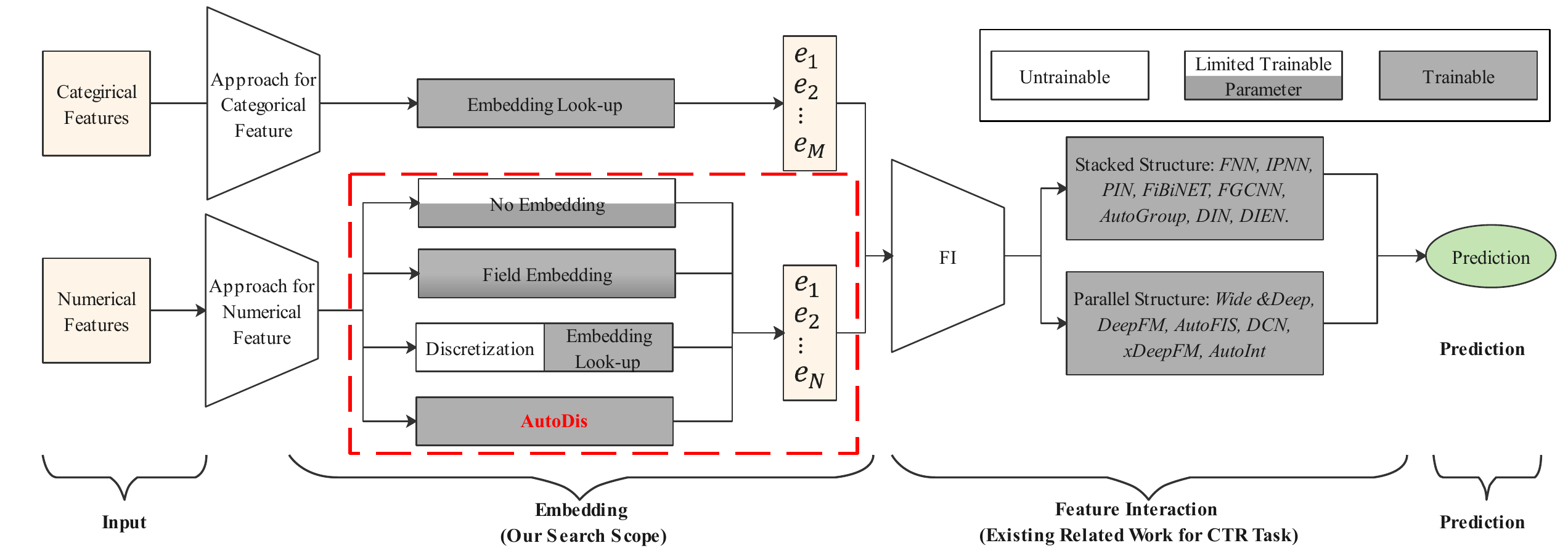}
    \caption{\small{Embedding \& Feature Interaction Architecture for CTR prediction.}}

    \label{fig:Embedding_FI}
\end{figure*}
As shown in Figure~\ref{fig:Embedding_FI}, most of the existing deep CTR models follow an \textit{Embedding \& Feature Interaction (FI)} paradigm.
Due to the significant importance for feature interaction in CTR prediction, the majority works focus on designing network architectures in \textit{FI module} to better capture explicit or implicit feature interactions, \textit{e.g.}, wide component in Wide \& Deep~\cite{wide-deep}, CrossNet in DCN~\cite{dcn}, and attention mechanism in DIN~\cite{din}.
Although not well studied in the literature, \textit{Embedding module} is also a key factor to deep CTR models for two reasons: (1) Embedding module is the cornerstone of the subsequent FI module and directly affects the effectiveness of FI module~\cite{Fnn}; (2) The number of parameters in deep CTR models is heavily concentrated in Embedding module, which naturally has high impact on the prediction performance~\cite{NIS}. However, the Embedding module is overlooked by the research community, which motivates us to conduct in-depth research.

The Embedding module works in a look-up table manner generally and maps each feature of the input data in categorical fields\footnote{For clarify, we use ``field" to represent a class of features following~\cite{ffm,pin} and ``feature¡± to represent a certain value in a specific field.}, to the latent embedding space with learnable parameters~\cite{deepfm}. Considering an example instance (\texttt{Gender}=Male, \texttt{Day}=Tuesday, \texttt{Height}=175.6, \texttt{Age}=18), for categorical field \texttt{Gender}, features are mapped to embeddings trivially by assigning a unique embedding to each individual feature, which is widely used in existing works~\cite{deepfm,pin}.
Unfortunately, such categorization strategy cannot be utilized to handle numerical features, as there may be infinitely many feature values in a numerical field (\textit{e.g.}, \texttt{Height}).

In practical, existing representation approaches for numerical feature can be summarized into three categories (red dashed box in Figure~\ref{fig:Embedding_FI}): (1) \textit{No Embedding}: using the original feature values or its transformation directly without learning embeddings~\cite{wide-deep,dnn-youtube,DLRM}; (2) \textit{Field Embedding}: learning a single field embedding for each numerical field~\cite{deepfm,song2019autoint}; (3) \textit{Discretization}: transforming numerical features to categorical features through various heuristic discretization strategies and assigning embedding like categorization strategy~\cite{pin}. However, the former two categories may lead to poor performance due to the low capacity of representations. The last category is also sub-optimal because such heuristic-based discretization rules cannot be optimized with the ultimate goal of CTR models. Besides, hard discretization-based methods suffer from \emph{SBD} (\textbf{S}imilar value \textbf{B}ut \textbf{D}is-similar embedding) and \emph{DBS} (\textbf{D}is-similar value \textbf{B}ut \textbf{S}ame embedding) problems whose details will be discussed in Session \ref{sec:pre}.

To address the limitations of existing approaches, we propose an \underline{auto}matic end-to-end embedding learning framework for numerical features based on soft \underline{dis}cretization (AutoDis). 
AutoDis consists of three core modules: \emph{meta-embeddings}, \emph{automatic discretization} and \emph{aggregation} to enable \textbf{\textit{high model capacity}}, \textbf{\textit{end-to-end training}} and \textbf{\textit{unique representation}} properties.
Specifically, we elaborately design a set of meta-embeddings for each numerical field, which are shared among all the intra-field feature values and learn global knowledge from the perspective of field with a manageable number of embedding parameters.
Then a differentiable automatic discretization module is leveraged to perform soft discretization and capture correlations between each numerical feature and field-specific meta-embeddings. Finally, an aggregation function is performed to learn unique \textit{Continuous-But-Different} representations.
To the best of our knowledge, AutoDis is the first end-to-end \emph{soft discretization} framework for numerical feature embedding that can be optimized jointly with the ultimate goal of deep CTR models.
We summarize the main contributions as follows.
\begin{itemize}[leftmargin=*]
    \item We propose AutoDis, a pluggable embedding learning framework for numerical features, which has high model capacity and is able to generate unique representations with a manageable number of parameters in an end-to-end manner.
    \item In AutoDis, we design meta-embeddings for each numerical field to learn global shared knowledge. Besides, a differentiable automatic discretization is used to capture the correlations between the numerical features and meta-embeddings, while an aggregation process is performed to learn a unique \textit{Continuous-But-Different} representation for each feature.
    \item Comprehensive experiments are conducted on two public and an industrial datasets to demonstrate the superiority of AutoDis over the existing representation methods for numerical features. Moreover, AutoDis works compatibly with various popular deep CTR models, by improving their recommendation performance significantly. 
    Online A/B test in a mainstream advertising platform shows that AutoDis improves the commercial baseline by 2.1\% and 2.7\% in terms of CTR and eCPM.
\end{itemize}

\section{Preliminary}\label{sec:pre}
In this section, we delve into the feature embedding learning procedure in CTR prediction.
Suppose a dataset for training CTR models consists of $Q$ instances $(\mathbf{x},y)$, where $y$ is label and $\mathbf{x}$ is a multi-field data record including $M$ categorical fields and $N$ numerical fields:
\begin{equation}
\small
\label{instance}
    \mathbf{x} = [\underbrace{\mathbf{x}_{1},\mathbf{x}_{2},\dots,\mathbf{x}_{M}}_{\text{one-hot vectors}};\underbrace{x_{1},x_{2},\dots,x_{N}}_{\text{scalars}}],
\end{equation}
where $\mathbf{x}_{i}$ is the one-hot vector of the feature value in the $i$-th categorical field and $x_{j}$ is the scalar value of the $j$-th numerical field.

For the $i$-th categorical field, the feature embedding can be obtained by embedding look-up operation:
\begin{equation}
\small
\label{categorical_feature}
    \mathbf{e}_{i} = \mathbf{E}_{i}\cdot\mathbf{x}_{{i}},
\end{equation}
where $\mathbf{E}_{i} \in \mathbb{R}^{v_{i}\times d}$ is the embedding matrix for $i$-th field, $v_{i}$ and $d$ is the vocabulary size and embedding size. Therefore, the representations for categorical fields are constructed as $[\mathbf{e}_{1},\mathbf{e}_{2},\dots,\mathbf{e}_{M}] \in \mathbb{R}^{M \times d}$.
For numerical fields, existing representation approaches can be summarized into three categories: \textit{No Embedding}, \textit{Field Embedding} and \textit{Discretization}.

\subsection{Category 1: No Embedding}
This category uses the original values or the transformations directly without learning embedding.
For instance, Wide \& Deep~\cite{wide-deep} of Google Play and DMT~\cite{DMT} of JD.com use the original and normalized numerical features respectively.
Besides, YouTube DNN~\cite{dnn-youtube} utilizes several transformations (\textit{i.e.}, square and square root) of normalized feature value $\tilde{x}_{j}$:
\begin{equation}
    \mathbf{e}_{{YouTube}} = [\tilde{x}_{1}^2,\tilde{x}_{1},\sqrt{\tilde{x}_{1}},\tilde{x}_{2}^2,\tilde{x}_{2},\sqrt{\tilde{x}_{2}},\dots,\tilde{x}_{N}^2,\tilde{x}_{N},\sqrt{\tilde{x}_{N}}],
\end{equation}
where $\mathbf{e}_{{YouTube}} \in \mathbb{R}^{3N}$.
In DLRM~\cite{DLRM} of Facebook, they use a multi-layer perception to model all numerical features:
\begin{equation}
\mathbf{e}_{{DLRM}}=[DNN([x_{1}, x_{2},\dots,x_{N}])],
\end{equation}
where the structure of $DNN$ is 512-256-$d$ and $\mathbf{e}_{{DLRM}} \in \mathbb{R}^{d}$.
Intuitively, these \textit{No Embedding} methods are difficult to capture informative knowledge of numerical fields due to their low capacity.

\subsection{Category 2: Field Embedding}
The commonly used method in academia is \textit{Field Embedding}~\cite{deepfm,song2019autoint}, where all the numerical features in the same field share a uniform field embedding and then multiply this field embedding with their feature values:
\begin{equation}
    \mathbf{e}_{FE} = [x_{1}\cdot \mathbf{e}_{1},x_{2}\cdot \mathbf{e}_{2},\dots,x_{N}\cdot \mathbf{e}_{N}],
\end{equation}
where $\mathbf{e}_j\in \mathbb{R}^{d}$ is the uniform embedding vector for the $j$-th numerical feature field and $\mathbf{e}_{{FE}} \in \mathbb{R}^{N\times d}$. However, the representation capacity of \textit{Field Embedding} is limited due to the single shared field-specific embedding as well as the linear scaling relationship between different intra-field features.


\subsection{Category 3: Discretization}
\label{dis}
A popular method in industrial recommender systems to handle numerical features is \textit{Discretization}, \textit{i.e.}, transforming numerical features to categorical features. For a feature $x_j$ in the $j$-th numerical field, the feature embedding $\mathbf{e}_j$ can be obtained by a two-stage operation: discretization and embedding look-up.
\begin{equation}
    \mathbf{e}_j = \mathbf{E}_j\cdot d_j(x_j),
\end{equation}
where $\mathbf{E}_j\in \mathbb{R}^{H_j\times d}$ is the embedding matrix for $j$-th field, $H_j$ is the number of buckets after discretization. Function $d_j( \cdot )$ is the manually-designed discretization method for the $j$-th numerical field, which project each feature value into one of the $H_j$ buckets. Specifically, there are three widely-used discretization approaches.

\noindent(1) \textbf{EDD/EFD} (Equal Distance/Frequency Discretization). EDD partitions the feature values into $H_j$ equal-width buckets. Suppose the range of features of the $j$-th field is $[x_j^{min},x_j^{max}]$, the interval width is defined as $w_j=(x_j^{max}-x_j^{min})/H_j$.
Accordingly, the discretized result $\widehat{x}_j$ is obtained via EDD discretization function $d_j^{EDD}( \cdot )$:
\begin{equation}
    \widehat{x}_j = d_j^{EDD}(x_j) = floor((x_j-x_j^{min})/w_j).
\end{equation}
Similarly, EFD divides the range $[x_j^{min}, x_j^{max}]$ to several buckets such that each bucket contains equal number of features.

\noindent(2) \textbf{LD} (Logarithm Discretization). The champion~\cite{ffm} of Criteo advertiser prediction in Kaggle Competition\footnote{http://https//www.kaggle.com/c/criteo-display-ad-challenge/discussion/10555\label{fn:kaggle}} utilizes the logarithm and floor operation to transform the numerical features to categorical form. The bucket $\widehat{x}_j$ after discretized is obtained by $d_j^{LD}( \cdot )$:
\begin{equation}
    \widehat{x}_j = d_j^{LD}(x_j) =  floor(log(x_j)^2).
\end{equation}

\noindent(3) \textbf{TD} (Tree-based Discretization). Apart from deep learning models, tree-based models (such as GBDT) are widely-used in recommendation because of their capability of handling numerical features effectively.
Therefore, many tree-based approaches are utilized to discretize the numerical features~\cite{fb_practical,DeepGBM,DT-FS}.

\begin{figure}[!t]
    \centering
    \setlength{\belowcaptionskip}{-0.5cm}
    \includegraphics[width=0.45\textwidth]{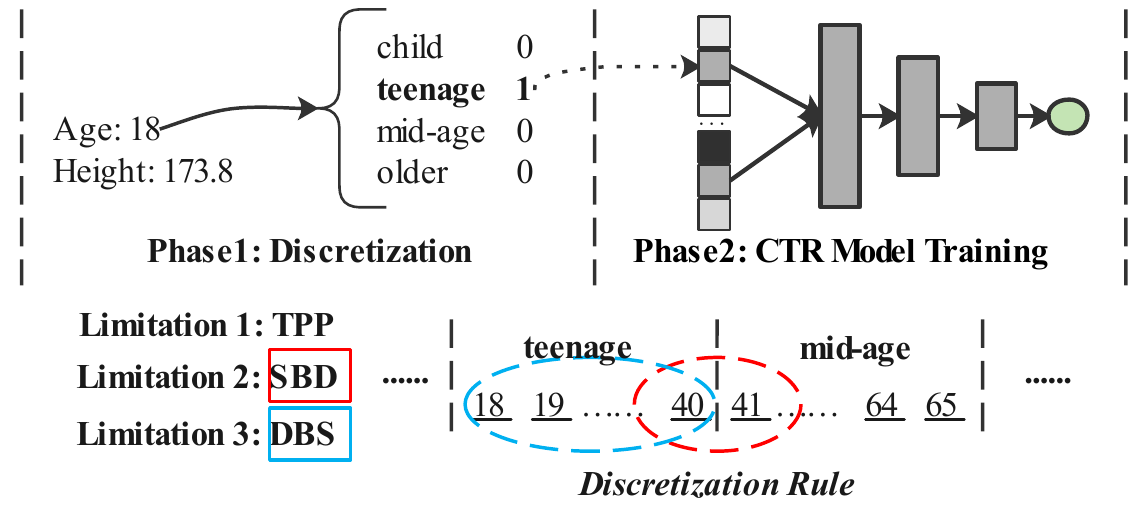}
    \caption{\small{Limitations of existing discretization approaches: (1) \emph{TPP} (\textbf{T}wo-\textbf{P}hase \textbf{P}roblem); (2) \emph{SBD} (\textbf{S}imilar value \textbf{B}ut \textbf{D}is-similar embedding); \emph{DBS} (\textbf{D}is-similar value \textbf{B}ut \textbf{S}ame embedding).}}

    \label{fig:limitations}
\end{figure}

Although \textit{Discretization} is widely-used in industry~\cite{pin,fb_practical}, they still suffer from three limitations (shown in Figure~\ref{fig:limitations}).
\begin{itemize}[leftmargin=*]
\item\emph{TPP} (\textbf{T}wo-\textbf{P}hase \textbf{P}roblem):  The discretization process is determined by either a heuristic rule or another model, so that it cannot be optimized together with the ultimate goal of the CTR prediction task, leading to sub-optimal performance.
\item\emph{SBD} (\textbf{S}imilar value \textbf{B}ut \textbf{D}is-similar embedding): These discretization strategies may separate similar features (boundary values) into two different buckets hence their afterwards embeddings are significantly different. For instance, a commonly used discretization for \texttt{Age} field is to determine [18,40] as teenage and [41,65] as mid-age, which results in significantly different embeddings for numerical value 40 and 41. 
\item\emph{DBS} (\textbf{D}is-similar value \textbf{B}ut \textbf{S}ame embedding): Existing discretization strategies may group significantly different elements into the same bucket, leading to indistinguishable embedding. Using the same example, numerical values between 18 and 40 are in the same bucket and therefore assigned with the same embedding. However, persons with 18-year-old and 40-year-old are likely with very different characteristics. \textit{Discretization}-based strategies cannot describe the continuity of numerical feature changes effectively.
 \end{itemize}

\subsection{Distinction and Relationship}

To summarize, a three-aspect comparison between AutoDis with existing representation approaches is presented in Table~\ref{tab:related}. We can observe that these methods are either difficult to capture informative knowledge because of low capacity or require offline expertise feature engineering, which might degrade the overall performance. Therefore, we propose AutoDis framework. To the best of our knowledge, it is the first numerical features embedding learning framework with \textbf{\textit{high model capacity}}, \textbf{\textit{end-to-end training}} and \textbf{\textit{unique representation}} properties preserved.


\begin{table}[!t]
\caption{\small{Comparison of approaches for numerical features representation learning.}}
\centering
\resizebox{0.9\columnwidth}{!}{%
\begin{tabular}{l|cccc}
\hline \hline
& Model & End-to-End & Unique Representation\\
& Capacity & Training & for Each feature\\ \hline
No Embedding & Limited & $\surd$ & $\times$  \\
Field Embedding & Limited  &$\surd$ & $\surd$ \\
Discretization & High &  $\times$ & $\times$ \\
AutoDis  & High & $\surd$ &$\surd$\\ \hline
\hline
\end{tabular}
}

\label{tab:related}
\end{table}

\section{Methodology}
\begin{figure}[!t]
    \centering
    \setlength{\belowcaptionskip}{-0.4cm}
 	\setlength{\abovecaptionskip}{-0cm}
    \includegraphics[width=0.45\textwidth]{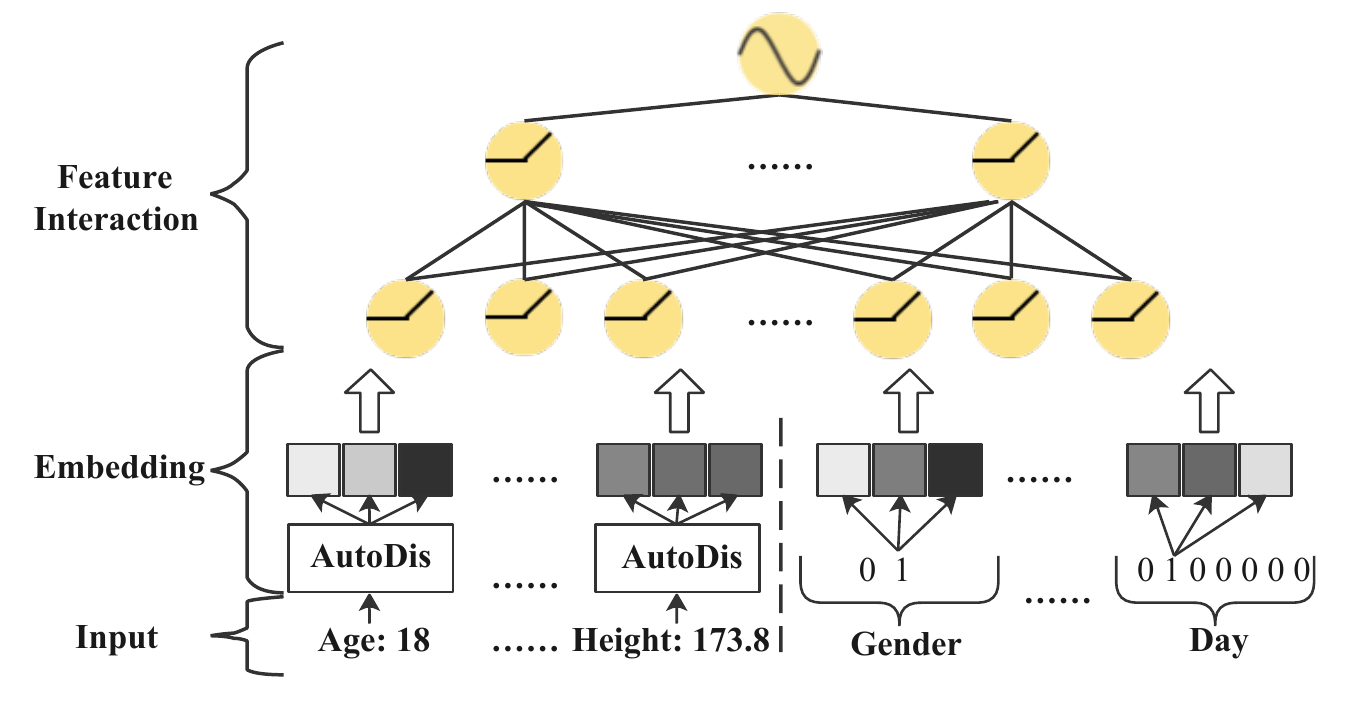}
    \caption{\small{AutoDis works as a numerical feature embedding learning framework compatible to the existing deep CTR models.}}
    \label{fig:architecture}
\end{figure}

To address the limitations of the existing approaches, we propose AutoDis, which automatically learns distinctive embeddings for numerical features in an end-to-end manner. As shown in Figure~\ref{fig:architecture}, AutoDis serves as a pluggable embedding framework for numerical features and is compatible with the existing deep CTR models.


\subsection{Framework Overview}
\begin{figure}[!t]
    \centering
    \setlength{\belowcaptionskip}{-0.4cm}
 	\setlength{\abovecaptionskip}{-0cm}
    \includegraphics[width=0.45\textwidth]{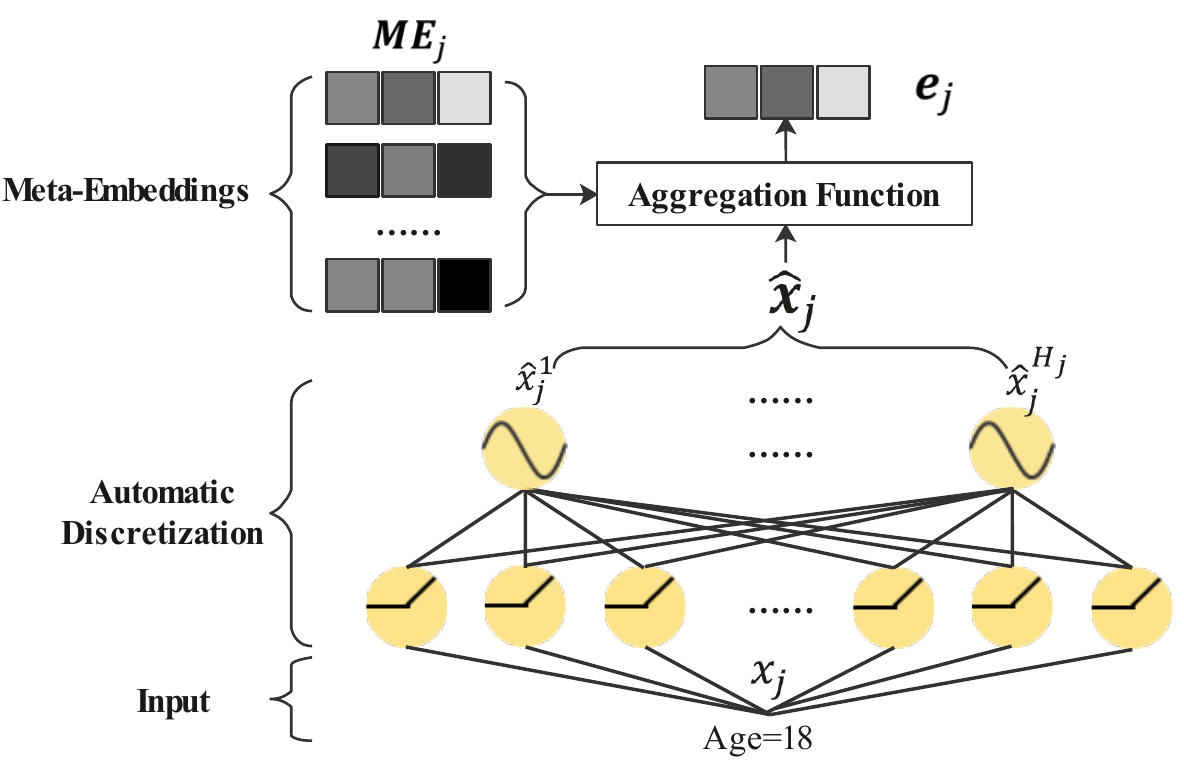}
    \caption{\small{The architecture of AutoDis framework.}}
    \label{fig:ELFNcrete}
\end{figure}

To preserve \textbf{\textit{high model capacity}}, \textbf{\textit{end-to-end training}} and \textbf{\textit{unique representation}} properties, we design three core modules: \emph{meta-embeddings}, \emph{automatic discretization} and \emph{aggregation} in AutoDis.
For the $j$-th numerical field, AutoDis can learn a unique representation for each numerical feature $x_{j}$ by:
\begin{equation}
\small
\label{numerical_feature}
    \mathbf{e}_{j} = f(d^{Auto}_j(x_{j}),\mathbf{ME}_{j}),
\end{equation}
where $\mathbf{ME}_{j}$ is the meta-embedding matrix for the $j$-th field, $d^{Auto}_j(\cdot)$ is the automatic discretization function and $f(\cdot)$ is the aggregation function. The architecture of AutoDis is presented in Figure~\ref{fig:ELFNcrete}.
Finally the embeddings of both categorical and numerical features are concatenated and fed into a deep CTR model for prediction:
\begin{equation}
\small
\label{deep_CTR}
    \hat y = \textit{CTR} (\underbrace{\mathbf{e}_{1},\mathbf{e}_{2},\dots,\mathbf{e}_{M}}_{\text{categorical embeddings}};\underbrace{\mathbf{e}_{1},\mathbf{e}_{2},\dots,\mathbf{e}_{N}}_{\text{numerical embeddings}}).
\end{equation}


\subsection{Meta-Embeddings}
A naive solution is to treat numerical features as categorical features and assign an independent embedding to each numerical feature. Obviously, this method has serious defects, such as explosively huge parameters and insufficient training for low-frequency features, which is unaffordable and inadvisable in practice. 
On the contrary, \textit{Field Embedding} method shares a single embedding among all the feature values within a field for saving parameters, degrading the model performance due to the low capacity.
To balance the model capacity and complexity, for the $j$-th field, we design a set of meta-embeddings $\mathbf{ME}_{j}\in \mathbb{R}^{H_j \times d}$ shared by features within this field, where $H_j$ is the number of meta-embeddings. Meta-embeddings are conductive to learn the intra-field shared global knowledge. Each meta-embedding can be viewed as a sub-space in the latent space for boosting expressiveness and capacity. By combining these meta-embeddings, the learned embedding is more informative than that of the \textit{Field Embedding} method, so that the \textbf{\textit{high model capacity}} can be well preserved. 
Moreover, the parameters needed are determined by $H_j$. Consequently, the model complexity is highly controllable, making our method scalable.


\subsection{Automatic Discretization}
\label{auto_dis}
To capture complicated correlations between a numerical feature value and the designed meta-embeddings, we meticulously propose a differentiable automatic discretization module $d^{Auto}_j(\cdot)$. Thanks to $d^{Auto}_j(\cdot)$, each numerical feature in the $j$-th field is discretized with $H_j$ buckets where each bucket embedding corresponds to a meta-embedding mentioned-above. 


Specifically, a two-layer neural network with skip-connection is leveraged to discretize the feature value $x_{j}$ to $H_j$ buckets by:

\begin{equation}
\label{mapping}
\begin{split}
& \mathbf{h}_{j} = \texttt{Leaky\_ReLU}(\mathbf{w}_{j} x_{j}), \\
& \mathbf{\widetilde{x}}_{j} = \mathbf{W}_j\mathbf{h}_{j} + \alpha\mathbf{h}_{j}, \\
\end{split}
\end{equation}
where $\mathbf{w}_{j}\in \mathbb{R}^{1\times H_j}$ and $\mathbf{W}_{j}\in \mathbb{R}^{H_j\times H_j}$ are the learnable parameters of the automatic discretization network for the $j$-th numerical feature field, activation function is $\texttt{Leaky\_ReLU}$~\cite{he2015delving} and $\alpha$ is the control factor of skip-connection. The projecting result is $\mathbf{\widetilde{x}}_{j} = [\widetilde{x}_{j}^1, \widetilde{x}_{j}^2, \dots, \widetilde{x}_{j}^{H_j}]$, where $\widetilde{x}_{j}^h$ is the projection output of feature value $x_{j}$ on the $h$-th bucket.
Finally, the correlations between feature value $x_{j}$ and each of the meta-embeddings $\mathbf{ME}_{j}$ is normalized via \texttt{Softmax} function, denoted as:
\begin{equation}
\small
\label{eq:softmax}
    \widehat{x}_{j}^h = \frac{e^{\frac{1}{\tau}\widetilde{x}_{j}^h}}{\sum_{l=1}^{H_j}e^{\frac{1}{\tau}\widetilde{x}_{j}^l}},
\end{equation}
where $\tau$ is the temperature coefficient to control discretization distribution. Therefore, the discretized result $\mathbf{\widehat{x}}_{j}\in \mathbb{R}^{H_j}$ is obtained through the automatic discretization function $d^{Auto}_j(\cdot)$:
\begin{equation}
\small
\label{eq:g_fun}
    \mathbf{\widehat{x}}_{j} = d^{Auto}_j(x_j) = [\widehat{x}_{j}^1, ..., \widehat{x}_{j}^h, ..., \widehat{x}_{j}^{H_j}].
\end{equation}

The discretized result $\mathbf{\widehat{x}}_{j}$ is a vector where each element $\widehat{x}_{j}^h$ denotes the probability that feature $x_{j}$ is discretized to the $h$-th bucket, which represents the correlation between feature value $x_{j}$ and the $h$-th meta-embedding (\textit{i.e.}, bucket embedding). This discretization method can be understood as \textit{soft discretization}. Compared with the \textit{hard discretization} mentioned in Section \ref{dis}, soft discretization does not discretize feature values into a single determinate bucket, so that the \emph{SBD} and \emph{DBS} issues can be well solved. Moreover, the differentiable soft discretization enables our AutoDis to achieve \textbf{\textit{end-to-end training}} and optimize with the ultimate goal.

It is worth noting that
when temperature coefficient in Eq.(\ref{eq:softmax}) $\tau \rightarrow \infty$, the discretization distribution in Eq.(\ref{eq:g_fun}) tends to be uniform and when $\tau \rightarrow 0$, the distribution tends to be one-hot. Therefore, temperature coefficient $\tau$ plays a significant role in the automatic discretization distribution. Besides, the feature distribution is diverse over different fields, so it is of great necessity to learn different $\tau$ for different features. Specifically, we propose a temperature coefficient adaptive network, which takes both global field statistics feature and local input feature into consideration, shown as:
\begin{equation}
\label{tao_net}
\tau_{x_{j}} = \texttt{Sigmoid}(\mathbf{W}_j^2\texttt{Leaky\_ReLU}(\mathbf{W}_j^1[\bar{\mathbf{n}}_j\Vert x_{j}])),
\end{equation}
where $\bar{\mathbf{n}}_j$ is the global statistics feature vector of the $j$-th field including sampled Cumulative Distribution Function (CDF) and mean value. $x_{j}$ is the local input feature, $\mathbf{W}_j^1$ and $\mathbf{W}_j^2$ are the weight parameters. To guide model training, the output range of $\tau_{x_{j}}$ is re-scale from (0, 1) to $(\tau-\epsilon, \tau+\epsilon)$, where $\tau$ is a global hyperparameter.



\subsection{Aggregation Function}
\label{agg_func}
After getting the correlations between a feature value and the meta-embeddings, the embeddings can be generated by applying aggregation operation $f(\cdot)$ over the meta-embeddings with the correlations as the selection probabilities.
The following aggregation functions $f(\cdot)$ are considered:

\begin{itemize}[leftmargin=*]
    \item \textbf{Max-Pooling} selects the most relevant meta-embedding with highest probability $\widehat{x}_{j}^h$:
    \begin{equation}
    \small
    \label{max}
    \mathbf{e}_{j} = \mathbf{ME}_{j}^k, \text{where } k = \arg max_{h\in\{1,2,\dots,H_j\}}\widehat{x}_{j}^h,
    \end{equation}
where $k$ is the index of the most relevant meta-embedding with highest probability $\widehat{x}_{j}^h$ and $\mathbf{ME}_{j}^k$ is the $k$-th meta-embedding in $\mathbf{ME}_{j}$.
However, this hard selection strategy degenerates AutoDis to a hard discretization method, resulting in aforementioned \emph{SBD} and \emph{DBS} problems.
\item \textbf{Top-K-Sum} sums up top-K meta-embeddings with highest correlation $\widehat{x}_{j}^h$:
\begin{equation}
\small
\label{topk}
    \mathbf{e}_{j} = \sum_{l=1}^K\mathbf{ME}_{j}^{k_l}, \text{where } k_l = \mathop{\arg top}_{l}\ _{h\in\{1,2,\dots,H_j\}}\widehat{x}_{j}^h,
\end{equation}
where $k_l$ is the index of meta-embedding with the $l$-th ($l\in[1, K]$) largest $\widehat{x}_{j}^h$, $\mathbf{ME}_{j}^{k_l}$ is the $k_l$-th meta-embedding in $\mathbf{ME}_{j}$. However, Top-K-Sum approach suffers from two limitations: (1) It can not fundamentally solve \emph{DBS} problem although the number of possible generated embeddings increases from $H_j$ to $C_{H_j}^K$ in comparison with Max-Pooling; (2) The learned embedding $\mathbf{e}_{j}$ does not consider the correlation values.
\item To take advantage of the entire set of meta-embeddings and their correlations to the feature value sufficiently, a \textbf{Weighted-Average} aggregation function is proposed as $\mathbf{e}_{j} = \sum_{h=1}^{H_j}\widehat{x}_{j}^h \cdot \mathbf{ME}_{j}^{h}$.
With this weighted aggregation strategy, relevant meta-embeddings contribute more to providing an informative embedding while irrelevant meta-embeddings are largely neglected. Besides, this strategy ensures that each feature learn \textbf{\textit{unique representation}}, and meanwhile, the learned embeddings are \emph{Continuous-But-Different}, meaning that the closer the feature values, the more similar the embeddings are.
\end{itemize}

\subsection{Training}
AutoDis is trained jointly with the ultimate goal of a concrete deep CTR model in an end-to-end manner. The loss function is the widely-used LogLoss with a regularization term:
\begin{equation}
\small
\label{loss}
    \mathcal{L} = -\frac{1}{Q}\sum_{i=1}^Q{y_i\log(\hat y_i)+(1-y_i)\log(1- \hat y_i)} + \lambda \Vert \Theta \Vert_2,
\end{equation}
where $y_i$ and $\hat y_i$ are the ground truth label and estimated value of the $i$-th instance, respectively. $Q$ is the total number of training instances and $\lambda$ is the $L_2$ regularization weight. $\Theta=\{\Theta_{Cat\_Emb}, \Theta_{AutoDis}, \\ \Theta_{CTR}\}$ are the parameters of feature embeddings in categorical fields, parameters of meta-embeddings and automatic discretization (\textit{i.e.}, Eq.(\ref{mapping}-\ref{tao_net})) in AutoDis, as well as deep CTR model parameters.

To stabilize the training process, we adopt the feature normalization technique to scale numerical feature values into $[0, 1]$ in the data pre-processing stage. A numerical feature value $x_{j}$ of the $j$-th numerical field is normalized as $x_{j} \gets \frac{x_{j} - x_{j}^{min}}{x_{j}^{max}-x_{j}^{min}}$.

\section{Experiments}
\subsection{Experimental Setting}
\subsubsection{Dataset and Evaluation Protocols}
To evaluate the effectiveness of our proposed numerical feature embedding learning framework AutoDis, we conduct extensive experiments on two popular benchmarks:~\emph{Criteo}, ~\emph{AutoML} and one industrial dataset. Table~\ref{data} summarizes the statistics of all the three datasets.
\begin{itemize}[leftmargin=*]
\item Criteo dataset is published in Criteo Display Advertising Challenge 2013\footnote{https://www.kaggle.com/c/criteo-display-ad-challenge} and is widely used in evaluating CTR models, which contains 13 numerical feature fields.
\item AutoML dataset is published from ¡°AutoML for Lifelong Machine Learning¡± Challenge in NeurIPS 2018\footnote{https://www.4paradigm.com/competition/nips2018}, which contains 23 numerical feature fields.
\item Industrial dataset is sampled and collected from a mainstream online advertising platform, with 41 numerical feature fields.
\end{itemize}

\begin{table}[!t]
\caption{\small{Statistics of evaluation datasets.}}
 \setlength{\belowcaptionskip}{-0.4cm}
 	\setlength{\abovecaptionskip}{-0cm}
\centering
\resizebox{.9\columnwidth}{!}{%
\begin{tabular}{c|c|c|c|c}
\hline \hline
Dataset & \#Num field & \#Cats field & \#Instances & Positive Ratio \\ \hline
Criteo & 13 & 26 & 45.8M & 25.6\% \\ \hline
AutoML & 23 & 51 & 4.69M & 5.8\% \\ \hline
Industrial & 41 & 44 & 8.75M & 3.3\% \\ \hline \hline
\end{tabular}
}
\label{data}
\end{table}

To fully evaluate the model performance, we leverage the most commonly-used evaluation metrics in CTR prediction, namely \textbf{AUC} and \textbf{LogLoss}. All the experiments are repeated 5 times by changing the random seeds. The two-tailed unpaired $t$-test is performed to detect
significant differences between AutoDis and the best baseline.

\subsubsection{Baselines and Hyper-parameters Settings}
To demonstrate the effectiveness of the proposed AutoDis, we compare AutoDis with three categories representation learning methods for numerical features: (1) \textit{No Embedding} (\textbf{YouTube}~\cite{dnn-youtube} \textbf{DLRM}~\cite{DLRM}); (2) \textit{Field Embedding} (\textbf{FE}~\cite{deepfm}); (3) \textit{Discretization} (\textbf{EDD}~\cite{pin}, \textbf{LD} as well as \textbf{TD}, namely DeepGBM~\cite{DeepGBM}).

Moreover, to validate the compatibility of AutoDis framework with various embedding-based deep CTR models, we apply AutoDis to six representative models:  \textbf{FNN}~\cite{Fnn}, \textbf{Wide \& Deep}~\cite{wide-deep}, \textbf{DeepFM}~\cite{deepfm}, \textbf{DCN}~\cite{dcn}, \textbf{IPNN}~\cite{pin} and \textbf{xDeepFM}~\cite{xdeepfm}.

Our implementation is publicly available in MindSpore. We optimize all the models with mini-batch Adam~\cite{adam}, where the learning rate is searched from
$\{$10e-6, 5e-5, $\dots$, 10e-2$\}$. Besides, the embedding size is set to 80 and 70 in Criteo and AutoML dataset. Specifically, the hidden layers in deep CTR models are fix to 1024-512-256-128 by default and the explicit feature interactions in DCN and xDeepFM (namely, CrossNet and CIN) are set to 3 layers. The weight of $L_2$ regularization is tuned in $\{$10e-6, 5e-5, $\dots$, 10e-3$\}$.
For AutoDis, the number of meta-embeddings in each numerical field is 20 and 40 in Criteo and AutoML. The skip-connection control factor is searched in $\{0, 0.1,\dots, 1\}$ and the number of neurons in temperature coefficient adaptive network is set to 64.

\subsection{Overall Performance}
In this section, we first compare the performance of different representation learning methods for numerical features. Then, we evaluate the compatibility of AutoDis by showing the improvement over six popular deep CTR models. 

\subsubsection{Compared with Other Representation Learning Methods}
We perform different numerical feature representation learning methods on three datasets and choose DeepFM as the deep CTR model. The performance comparison is presented in Table~\ref{results}, from which, we have the following observations:
\begin{itemize}[leftmargin=*]
    \item The baselines perform inconsistently in different datasets due to discrepant characteristics of numerical features. Comparatively speaking, AutoDis outperforms all the baselines on all the datasets consistently by a significant margin, demonstrating the superiority and robustness to different numerical features.

    \item \textit{No Embedding} and \textit{Field Embedding} methods perform worse than \textit{Discretization} and AutoDis. \textit{No Embedding} methods achieve better results on industrial dataset than on public datasets, while \textit{Field Embedding} method performs the opposite.
    These two category methods suffer from low capacity and limited expressiveness.
    On the one hand, \textit{No Embedding} in YouTube is parameter-free and DLRM leverages a DNN to obtain a mixed representation of all the numerical fields, ignoring the inter-field distinction, which limits the performance significantly.
    On the other hand, for \textit{Field Embedding}, all the feature values in the same field share a single embedding and the learned embeddings are the linear scaling of the shared embedding with the feature values. Such sharing and linear scaling strategy limits the performance fatally.
    \item In comparison with the three existing representative \textit{Discretization} methods that transform the numerical features into categorical form, AutoDis improves AUC over the best baseline by 0.17\%, 0.23\%, 0.22\%, respectively. Although \textit{Discretization} methods achieve better performance compared with the previous two categories, they cannot be optimized with the CTR models, and also suffer from \textit{SBD} and \textit{DBS} problems. The heuristic rule-based hard discretization strategies result in step-wise ``unsmooth'' embeddings.
    On the contrary, AutoDis overcomes these disadvantages by sharing a set of meta-embeddings for each numerical field and leveraging differentiable soft discretization and aggregation strategy to learn informative \textit{Continuous-But-Different} representations. 
\end{itemize}

\begin{table}[t]
\caption{\small{The overall performance comparison. Boldface denotes the highest score and underline indicates the best result of the baselines. $\star$ represents significance level $p$-value $<0.05$ of comparing
AutoDis with the best baselines.}}
 \setlength{\belowcaptionskip}{-0.4cm}
 	\setlength{\abovecaptionskip}{-0cm}
\resizebox{\linewidth}{!}{
\small
\begin{tabular}{l|c|c|c|c|c|c|c}
\hline\hline
\multirow{2}{*}{Category}
&
\multirow{2}{*}{Method} &
\multicolumn{2}{c|}{Criteo} & %
\multicolumn{2}{c|}{AutoML} &
\multicolumn{2}{c}{Industrial}\\
\cline{3-8}
 && AUC & LogLoss & AUC & LogLoss & AUC & LogLoss \\ \hline
\multirow{2}{*}{\textit{No Embedding}}&YouTube  &0.8104&0.4437&0.7382&0.1928&0.7259&0.1369 \\
&DLRM  &0.8114&0.4404&0.7525&0.1900&0.7253&0.1369\\ \hline
\textit{Field Embedding} & FE & 0.8107 & 0.4412 & 0.7523 & 0.1899 & 0.7248&0.1369 \\ \hline
\multirow{3}{*}{\textit{Discretization}}&EDD & 0.8125 & 0.4399 &\underline{0.7545}  &\underline{0.1898}  &0.7251&0.1371 \\
&LD & \underline{0.8138} & \underline{0.4388} & 0.7527 & 0.1899 & \underline{0.7265}& \underline{0.1368} \\
&TD  & 0.8130 &0.4392  & 0.7531  & 0.1899 & 0.7262 & 0.1369 \\ \hline
\multicolumn{2}{c|}{AutoDis} & \textbf{0.8152}$^\star$ &  \textbf{0.4370}$^\star$ & \textbf{0.7562}$^\star$ & \textbf{0.1892}$^\star$ &\textbf{0.7281}$^\star$ & \textbf{0.1366}$^\star$ \\ \hline
\multicolumn{2}{c|}{\% Improv.} & \textbf{0.17\%} & \textbf{0.41\%} & \textbf{0.23\%} & \textbf{0.32\%} & \textbf{0.22\%} & \textbf{0.15\%} \\ \hline \hline
\end{tabular}
}
\label{results}
\end{table}

\subsubsection{Compatibility with Different CTR Models}
AutoDis is a universal framework and can be regarded as plug-in component for improving various deep CTR models performance. To demonstrate its compatibility, in this section, we conduct extensive experiments by applying AutoDis on a series of popular models. From Table~\ref{results-models} we can observe that, compared with the \textit{Field Embedding} representation method, AutoDis framework significantly improves the prediction performance of these models. The numerical feature discretization and embedding learning procedure is optimized with the ultimate goal of these CTR models, so that the informative representations can be obtained and the performance can be improved.
In fact, a small improvement in AUC is likely to yield a significant online CTR increase and huge profits ~\cite{wide-deep,pin}.  


\begin{table}[t]

\caption{\small{Compatibility Study of AutoDis. \textit{Field Embedding} method is performed as the baseline to compare with AutoDis. }}

    \centering
\resizebox{\linewidth}{!}{
\small
\begin{tabular}{l|c|c|c|c|c|c}
\hline\hline
&
\multicolumn{2}{c|}{Criteo} & %
    \multicolumn{2}{c|}{AutoML} &
    \multicolumn{2}{c}{Industrial}\\
\cline{2-7}
 & AUC & LogLoss & AUC & LogLoss & AUC & LogLoss \\ \hline
FNN & 0.8059 & 0.4456 & 0.7383 & 0.1926  & 0.7271 &0.1367  \\
FNN-AutoDis & 0.8092$^\star$ &  0.4426$^\star$ & 0.7453$^\star$ & 0.1907$^\star$  & 0.7289$^\star$ & 0.1365$^\star$  \\ \hline
Wide\&Deep & 0.8097 & 0.4419 & 0.7407 & 0.1923  & 0.7275 & 0.1366  \\
Wide\&Deep-AutoDis & 0.8121$^\star$ & 0.4390$^\star$ & 0.7445$^\star$ & 0.1914$^\star$  & 0.7288$^\star$ & 0.1365$^\star$  \\ \hline
DeepFM & 0.8107 & 0.4412 & 0.7523 & 0.1899  & 0.7248 & 0.1369  \\
DeepFM-AutoDis & 0.8152$^\star$ & 0.4370$^\star$ & 0.7562$^\star$ & 0.1892$^\star$  & 0.7281$^\star$ & 0.1366$^\star$  \\ \hline
DCN & 0.8091 & 0.4425 & 0.7489 & 0.1909  & 0.7262 &0.1369  \\
DCN-AutoDis & 0.8132$^\star$ & 0.4395$^\star$ & 0.7511$^\star$ & 0.1898$^\star$  & 0.7293$^\star$ &0.1364$^\star$  \\ \hline
IPNN & 0.8101 &0.4415 & 0.7519 & 0.1896  & 0.7269 & 0.1366 \\
IPNN-AutoDis & 0.8136$^\star$& 0.4383$^\star$ & 0.7541$^\star$& 0.1894$^\star$ & 0.7286$^\star$ & 0.1365$^\star$ \\ \hline
xDeepFM &0.8103 &0.4414 & 0.7508&0.1903 & 0.7275&0.1367 \\
xDeepFM-AutoDis &0.8141$^\star$ &0.4381$^\star$ &0.7529$^\star$ &0.1897$^\star$ & 0.7289$^\star$ & 0.1365$^\star$ \\ \hline \hline

\end{tabular}
}
\label{results-models}
\end{table}

\subsection{Deployment \& Online A/B Testing}
Online experiments were conducted in a mainstream advertising platform for a weak to verify the superior performance of AutoDis, where millions of daily active users interact with ads and generate tens of millions of user log events. In the online serving system, thousands of candidate ads that are most popular or relevant to users are retrieved from the universal ads pool and then ranked by a highly-optimized deep CTR model before presenting to users. To guarantee user experience, the overall latency of the above-mentioned candidate recall and ranking is required to be within a few milliseconds. The commercial model is deployed in a cluster, where each node is with 48 core Intel Xeon CPU E5-2670 (2.30GHZ), 400GB RAM as well as 2 NVIDIA TESLA V100 GPU cards.
For the control group, the numerical features are discretized via a variety of hybrid manually-designed rules (\textit{e.g.}, EDD, TD, \textit{etc.}). The experience group chooses AutoDis to discretize and learn embeddings automatically for all the numerical features. Deploying AutoDis to an existing CTR model is convenient and requires nearly no engineering work in the online serving system. 

AutoDis achieves an offline AUC 0.2\% improvement, and upgrades the online CTR and eCPM by 2.1\% and 2.7\% relative to the control group\footnote{The base model in control group serves the major proportion of users.} (statistically significant), which brings enormous commercial profits. 
Moreover, with AutoDis integrated, the existing numerical features no longer require any discretization rules. Furthermore, introducing more numerical features in the future will be more efficient, without exploring any handcrafted rules.


\subsection{Embedding Analysis}
To understand the \textit{Continuous-But-Different} embeddings learned via AutoDis more deeply with insights, we do further investigation in macro-analysis of embeddings and micro-analysis of soft discretization, respectively. 

\subsubsection{Macro-Analysis of Embeddings}
\begin{figure}
	\centering
	\setlength{\belowcaptionskip}{-0.3cm}
	\setlength{\abovecaptionskip}{0cm}
	\subfigure[\small{AutoDis embeddings}]{
		\label{fig:ELFN}
		\includegraphics[width=0.21\textwidth]{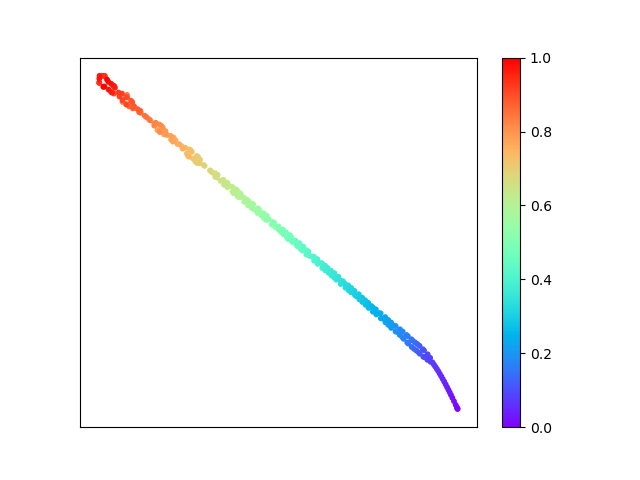}}
	\subfigure[\small{EDD embeddings}]{
		\label{fig:edd}
		\includegraphics[width=0.21\textwidth]{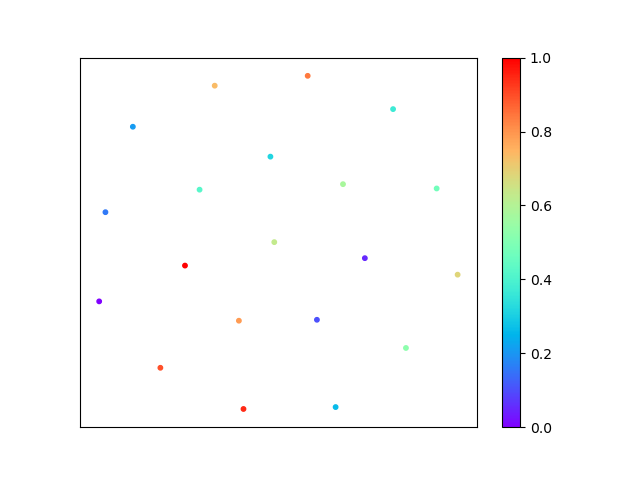}}
	\caption{\small{Visualization of the t-SNE transformed embeddings derived from AutoDis and EDD. The points in different colors reflect different feature values after normalization.}}
	\label{fig:embedding_visualization}
\end{figure}

Figure~\ref{fig:embedding_visualization} provides a visualization of the embeddings derived from DeepFM-AutoDis and DeepFM-EDD in the $3$-rd numerical field of the Criteo dataset. We random select 250 embeddings and project them onto a two dimensional space using t-SNE~\cite{van2008visualizing}. Nodes with similar colors have similar values. We can observe that AutoDis learns a unique embedding for each feature. Moreover, similar numerical features (with similar colors) are represented by close-correlated embeddings (with similar positions in two-dimensional space), which elaborates the \textit{Continuous-But-Different} property of the embeddings. However, EDD learns identical embedding for all the features within a bucket and completely different embeddings across different buckets, which leads to step-wise ``unsmooth'' embeddings and consequently results in inferior task performance.


\subsubsection{Micro-Analysis of Soft Discretization}
\begin{figure}
	\centering
	\setlength{\belowcaptionskip}{-0.4cm}
 	\setlength{\abovecaptionskip}{-0cm}
	\subfigure[\small{Feature value ``1" \textit{v.s.} Feature value ``2"}]{
		\label{fig:softmax1}
		\includegraphics[width=0.21\textwidth]{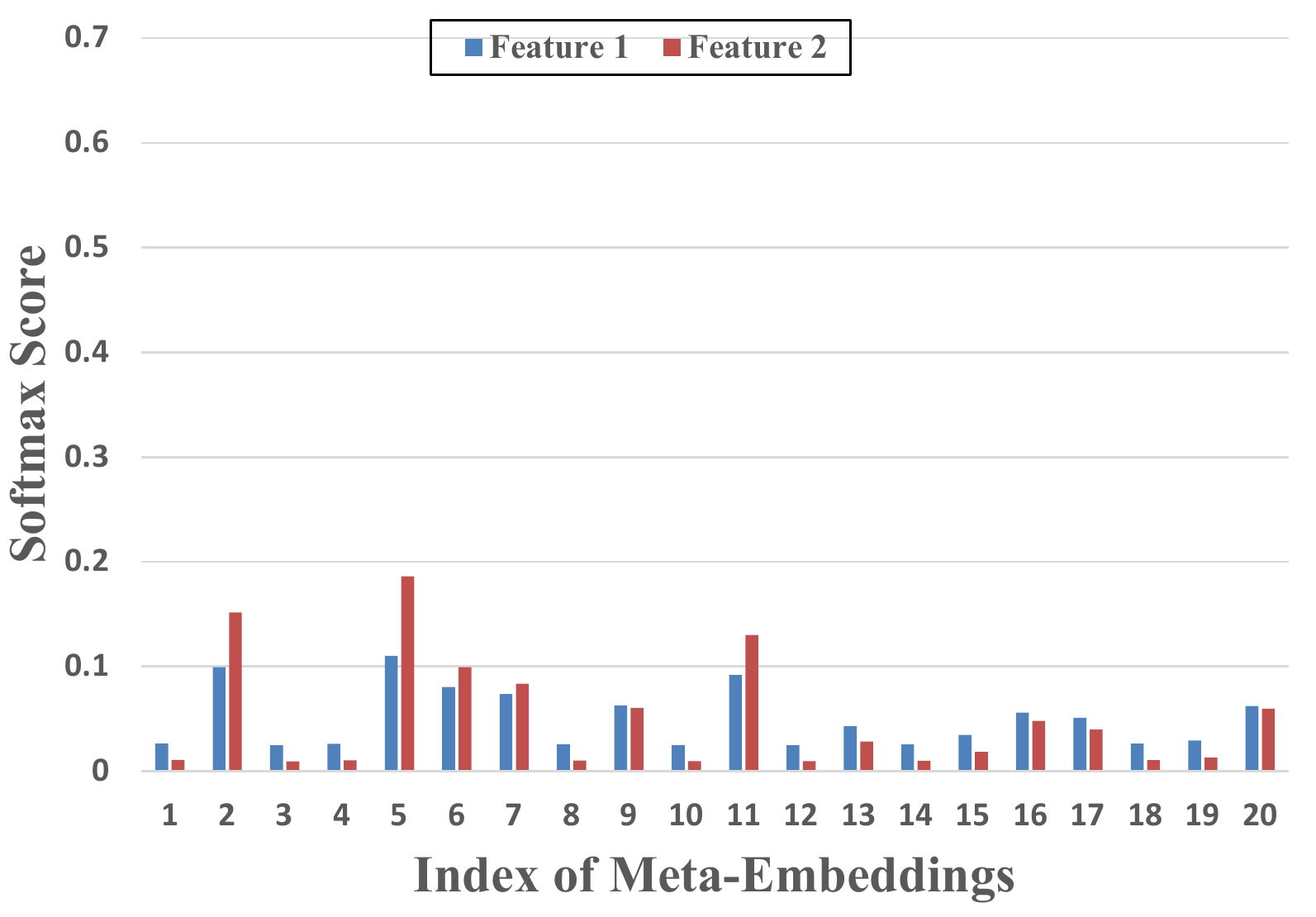}}
 	\hspace{2em}
	\subfigure[\small{Feature value ``1" \textit{v.s.} Feature value ``10"}]{
		\label{fig:softmax2}
		\includegraphics[width=0.21\textwidth]{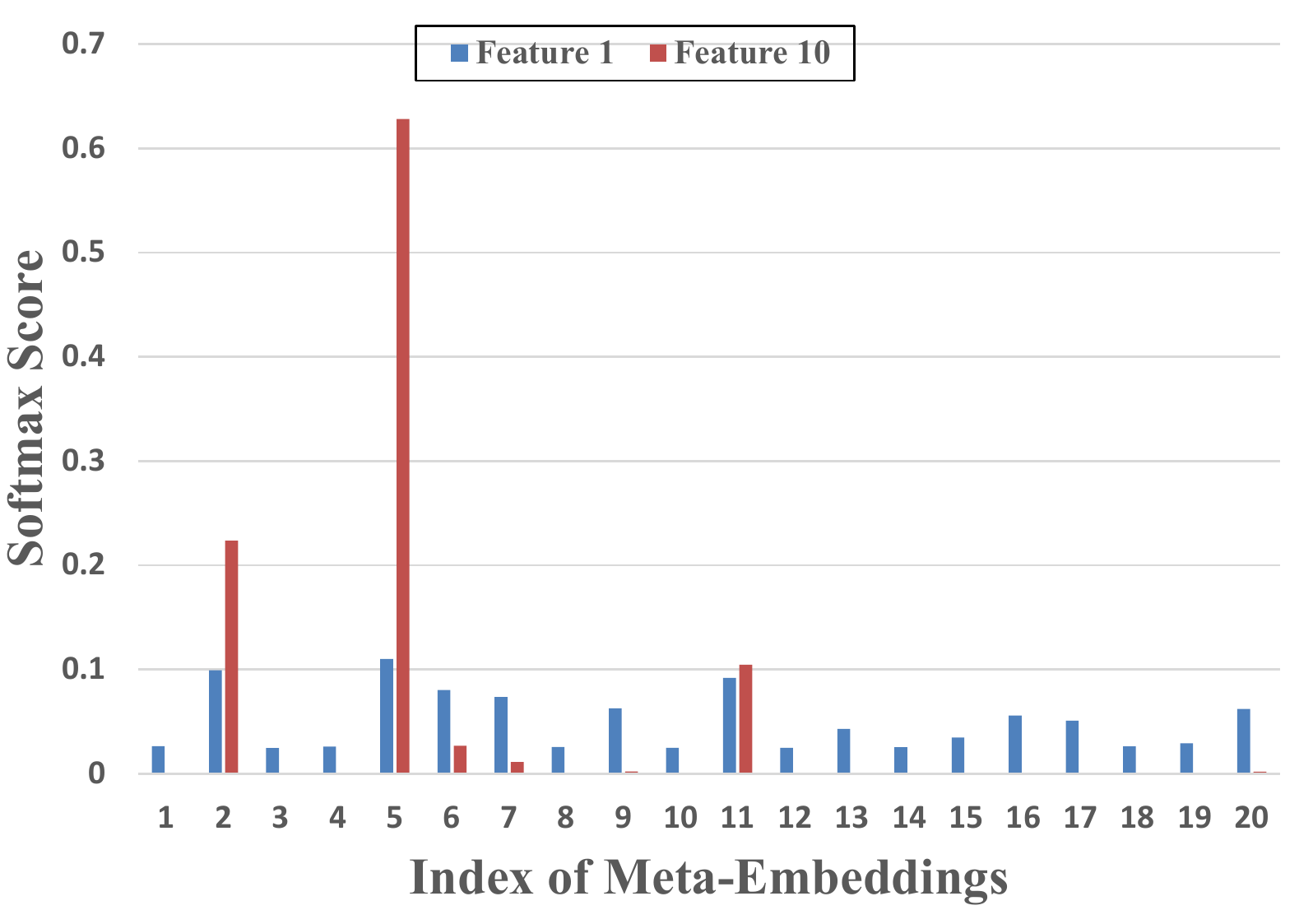}}
	\caption{\small{Soft discretization distribution in the $8$-th field of Criteo.}}
	\label{fig:softmax}
\end{figure}

In this section, we perform some micro-level analysis by investigating the Softmax distribution during the soft discretization process from DeepFM-AutoDis. We select an adjacent feature-pair (feature value ``1'' and ``2'') and a distant feature-pair (feature value ``1'' and ``10'') from the $8$-th numerical field of the Criteo dataset, and then visualize their discretization distributions in Figure~\ref{fig:softmax}. We can observe that, the adjacent feature-pair have similar Softmax distributions while distant feature-pair have diverse distributions. This characteristic is conducive to ensure that similar feature values can learn similar embeddings by AutoDis, so that the embedding continuity can be preserved.


\subsection{Numerical Fields Analysis}

\begin{figure}[!t]
	\centering
	\setlength{\belowcaptionskip}{-0.3cm}
	\setlength{\abovecaptionskip}{0cm}
	\subfigure[The original order]{
		\label{fig:fix_features}
		\includegraphics[width=0.21\textwidth]{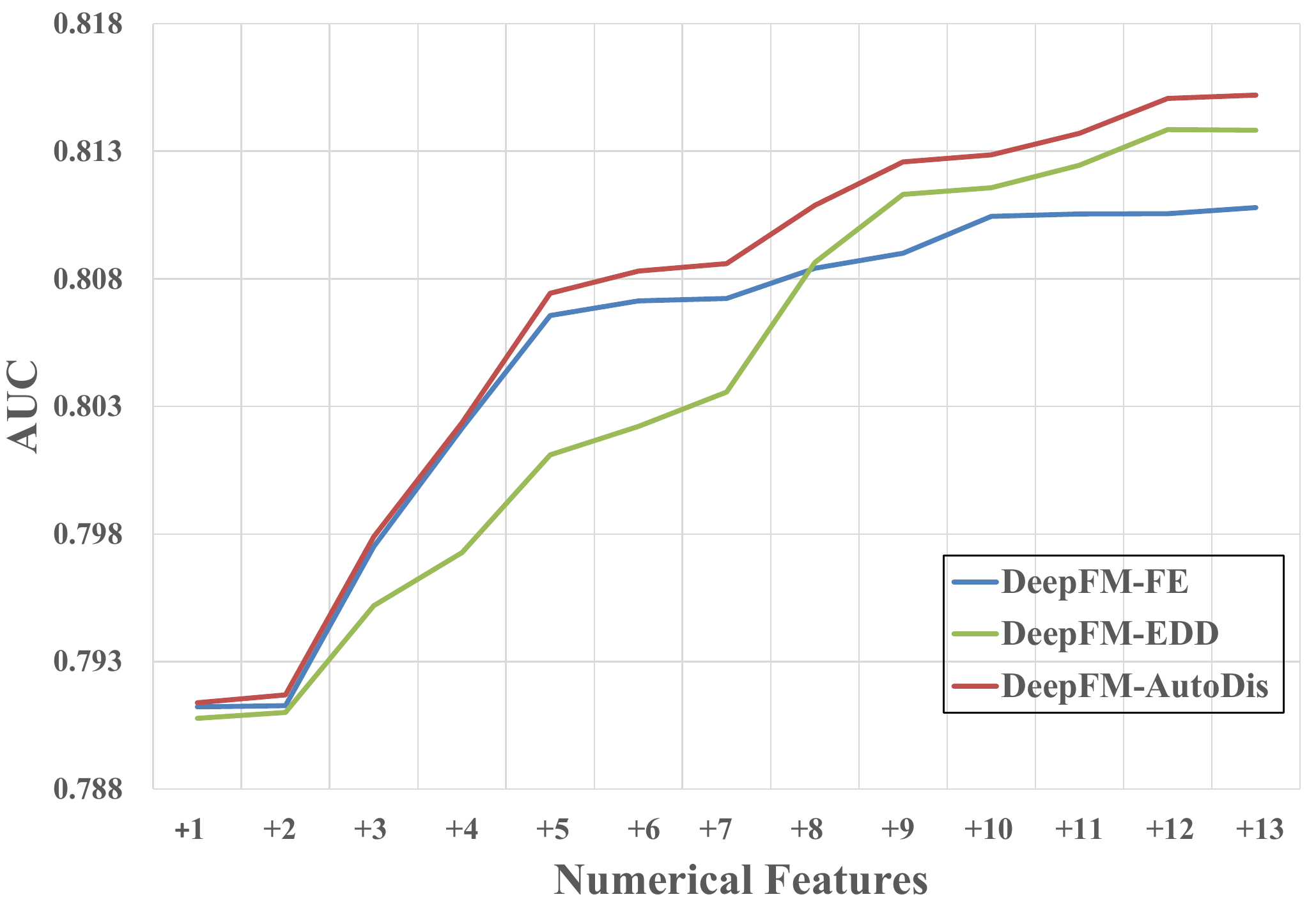}}
 	\hspace{1em}
	\subfigure[A random order]{
		\label{fig:random_features}
		\includegraphics[width=0.21\textwidth]{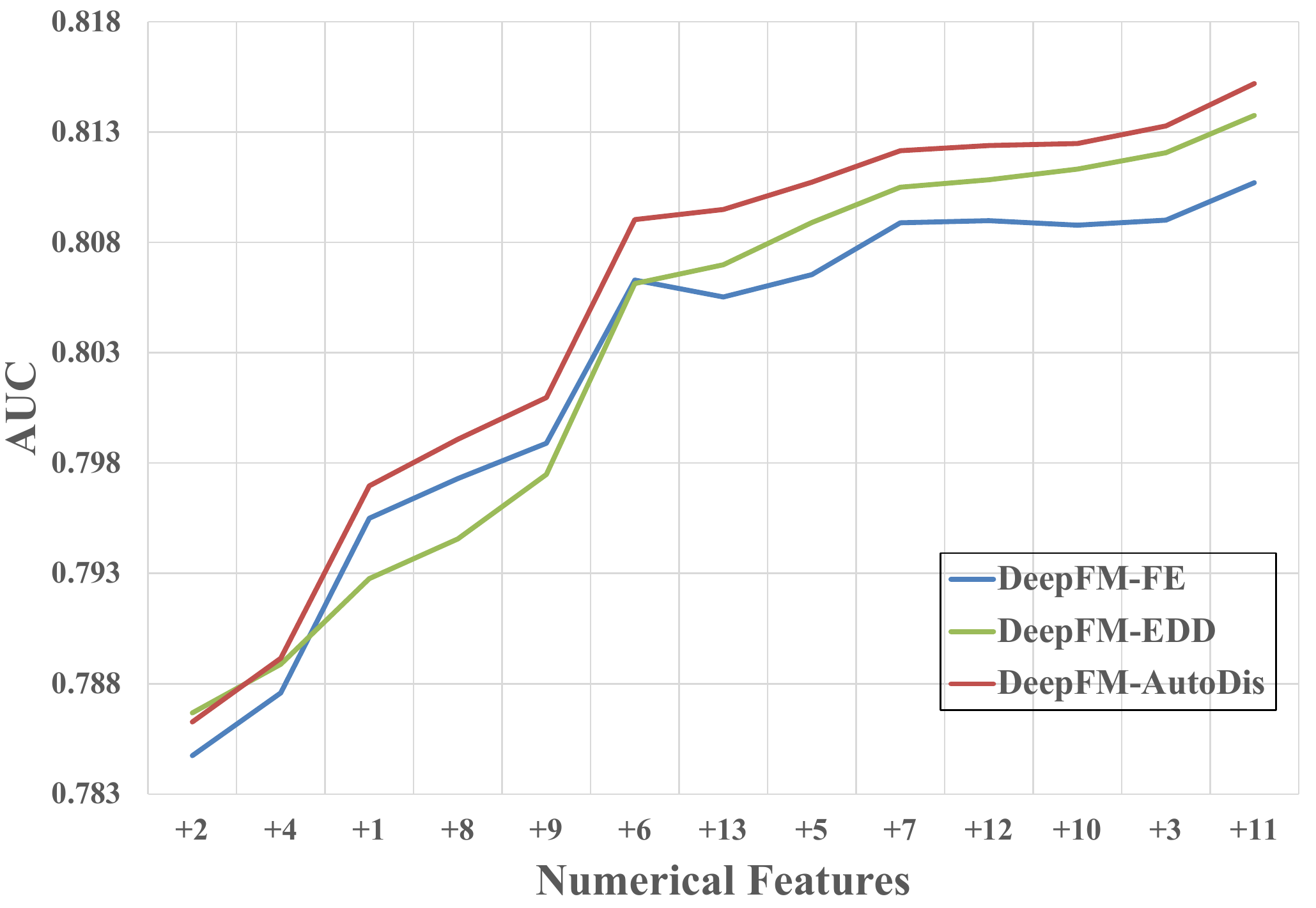}}
	\caption{\small{Improvements by introducing numerical fields one at a time by different orders.}}
	\label{fig:con_features}
\end{figure}

To evaluate the effect of DeepFM-AutoDis on each numerical field, we select all the 26 categorical fields as the basic features and add one of the 13 numerical fields at a time cumulatively in the Criteo dataset. Figure~\ref{fig:con_features} plots the prediction performance \textit{w.r.t} adding numerical fields in the original order of the dataset and in a random order. From Figure~\ref{fig:con_features}, we have the following observations: (1) AutoDis improves the performance even if there is only a single numerical field; (2) AutoDis has a cumulative improvement on multiple numerical fields; (3) Compared with the existing methods, the performance improvement achieved by AutoDis is more significant and stable.

\subsection{Model Complexity}

To quantitatively analyze the space and time complexity of our proposed AutoDis, the model parameters and batch inference time with the EDD discretization method on DeepFM are compared. Table~\ref{complexity} reports the results on Criteo dataset. We can observe that, compared with EDD, the increased amount of model parameters by AutoDis are negligible.
As online recommendation services usually have high requirements on latency, the computational cost during inference is important. We can find that AutoDis achieves acceptable inference time compared to EDD, demonstrating its computational efficiency.
Furthermore, AutoDis is an end-to-end framework, indicating that there is no extra efficiency overhead. However, other \textit{Discretization} methods need additional overhead, \textit{e.g.}, manual design and data preprocessing in EDD; another GBDT model training in TD.

\begin{table}[!t]
\caption{\small{Model complexity comparison.}}
\centering
\resizebox{.8\columnwidth}{!}{%
\begin{tabular}{c|c|c}
\hline \hline
Model & Parameters&Batch inference time \\ \hline
DeepFM-EDD & $\sim$2.009M &$\sim$3.48ms \\ \hline
DeepFM-AutoDis & $\sim$2.012M  & $\sim$3.63ms  \\ \hline
Relative Ratio & +0.15\% & +4.31\%\\ \hline
\hline
\end{tabular}
}
\label{complexity}
\end{table}

\subsection{Ablation Study and Hyper-Parameters Sensitivity}

\subsubsection{Aggregation Strategy}
\begin{figure}[!t]
	\centering
	\setlength{\belowcaptionskip}{-0.3cm}
	\setlength{\abovecaptionskip}{0cm}
	\subfigure[Criteo]{
		\label{fig:strategy_criteo}
		\includegraphics[width=0.21\textwidth]{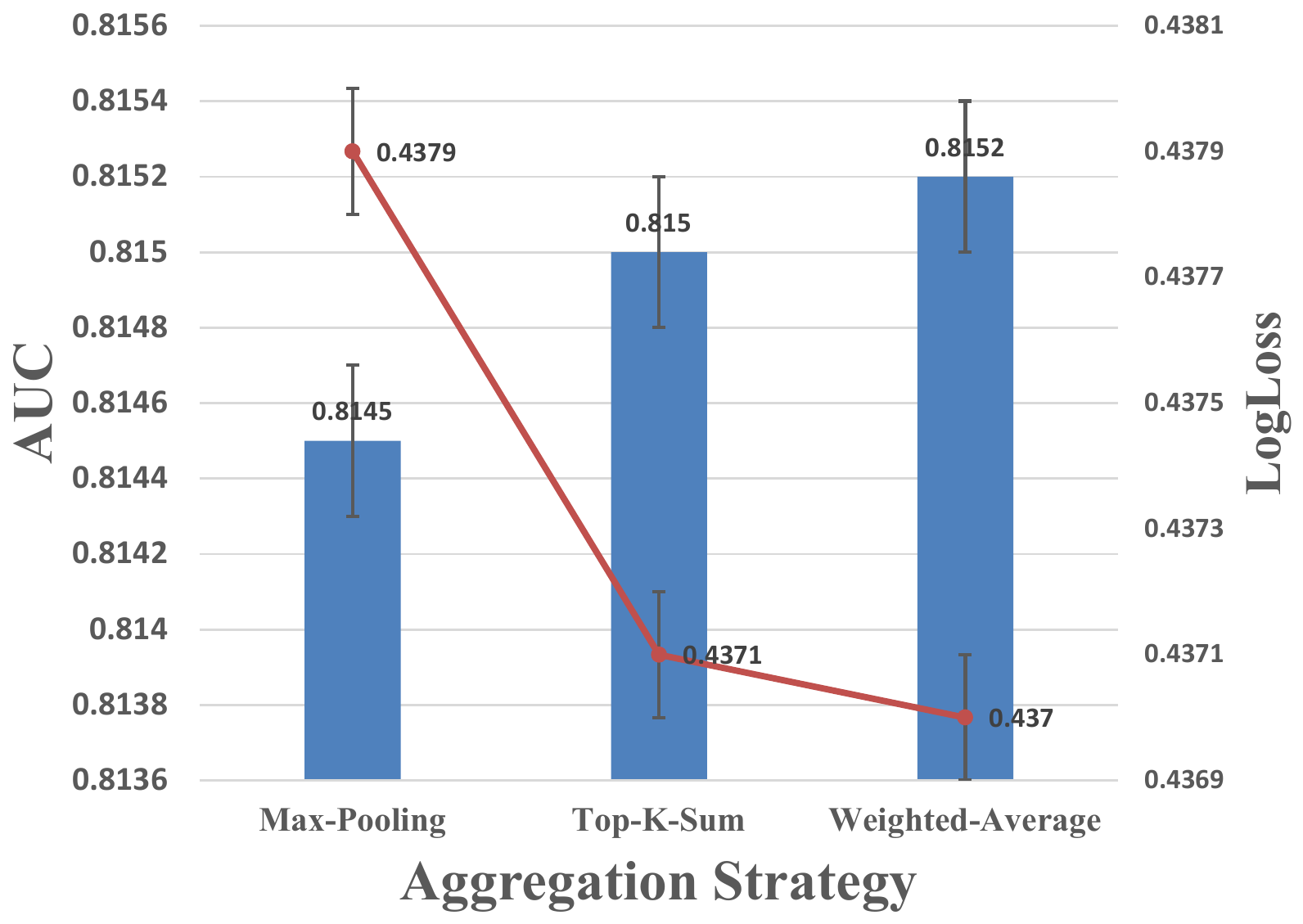}}
 	\hspace{1em}
	\subfigure[AutoML]{
		\label{fig:strategy_automl}
		\includegraphics[width=0.21\textwidth]{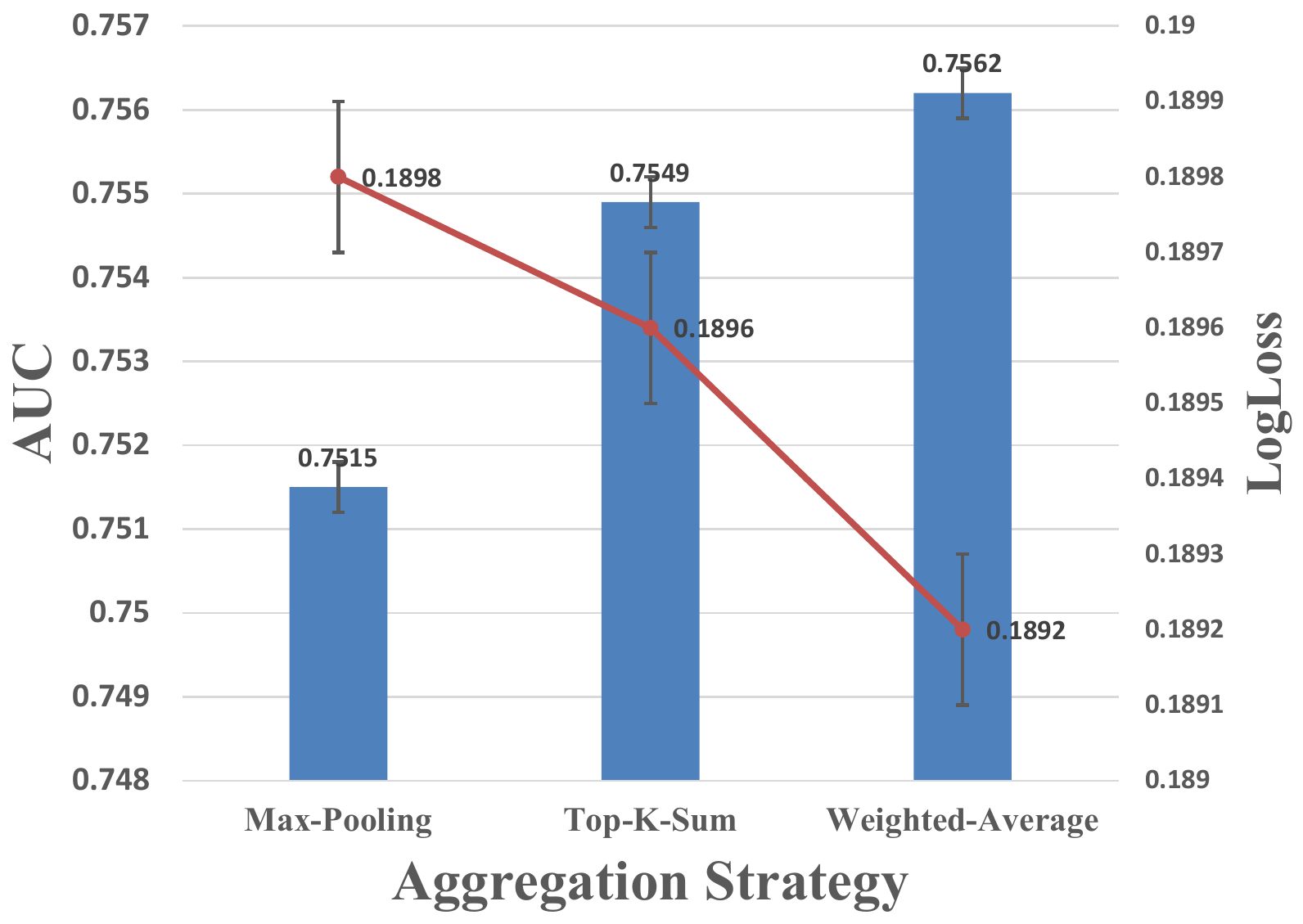}}
	\caption{\small{The effect of aggregation strategy.}}
	\label{fig:strategy}
\end{figure}

To verify the effect of different aggregation strategies in Section \ref{agg_func}, we perform ablation study. Figure~\ref{fig:strategy} summarizes the prediction performance of DeepFM-AutoDis with Max-Pooling, Top-K-Sum and Weighted-Average aggregation strategies. It can be concluded that Weighted-Average strategy achieves the best performance. The reason is that, Weighted-Average makes full use of the meta-embeddings and their corresponding correlations, and overcomes the \textit{DBS} and \textit{SBD} problems completely compared with other strategies.


\subsubsection{Temperature Coefficient}
\begin{figure}
	\centering
	\setlength{\belowcaptionskip}{-0.3cm}
	\setlength{\abovecaptionskip}{0cm}
	\subfigure[\small{Criteo}]{
		\label{fig:temp_criteo}
		\includegraphics[width=0.21\textwidth]{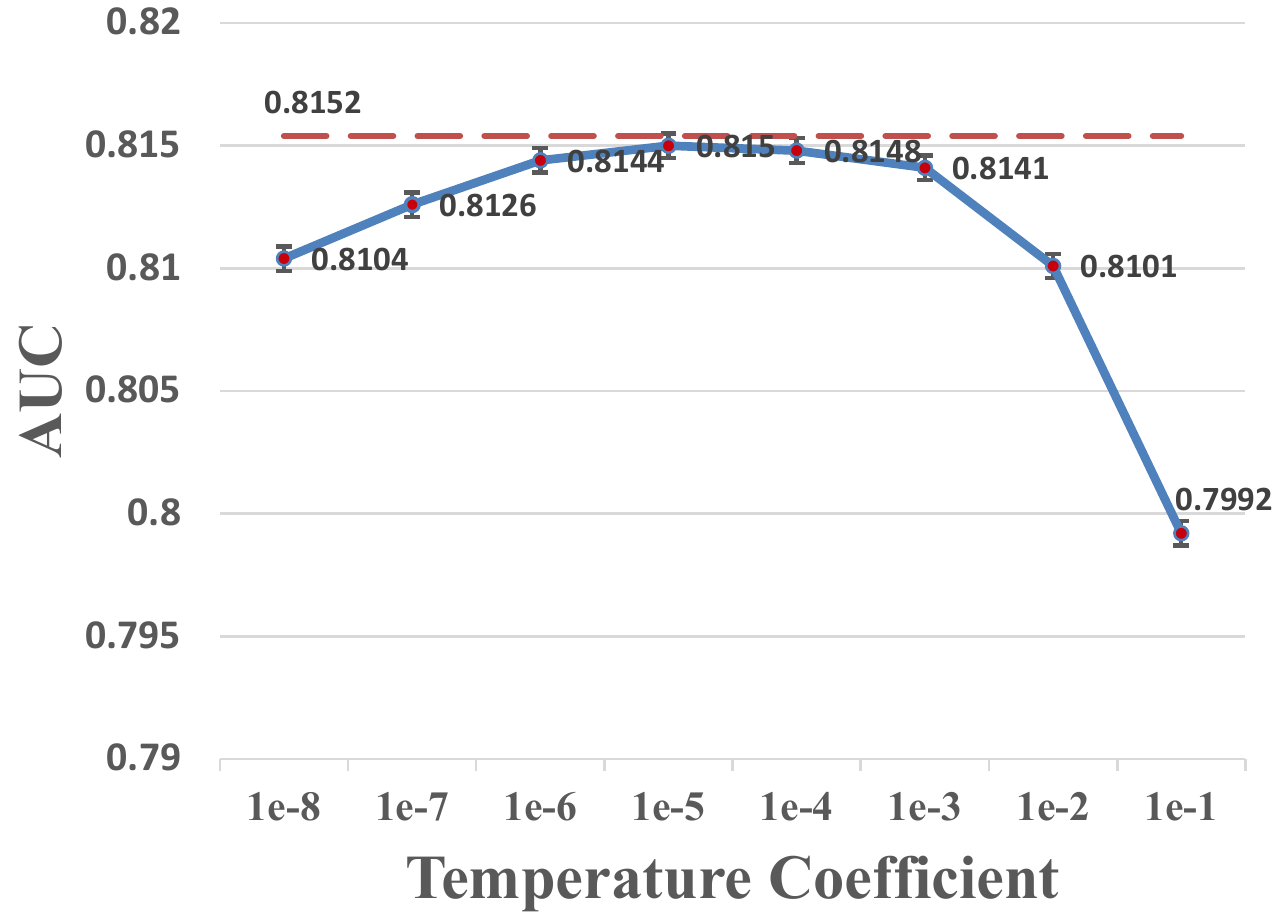}}
 	\hspace{1em}
	\subfigure[\small{AutoML}]{
		\label{fig:temp_automl}
		\includegraphics[width=0.21\textwidth]{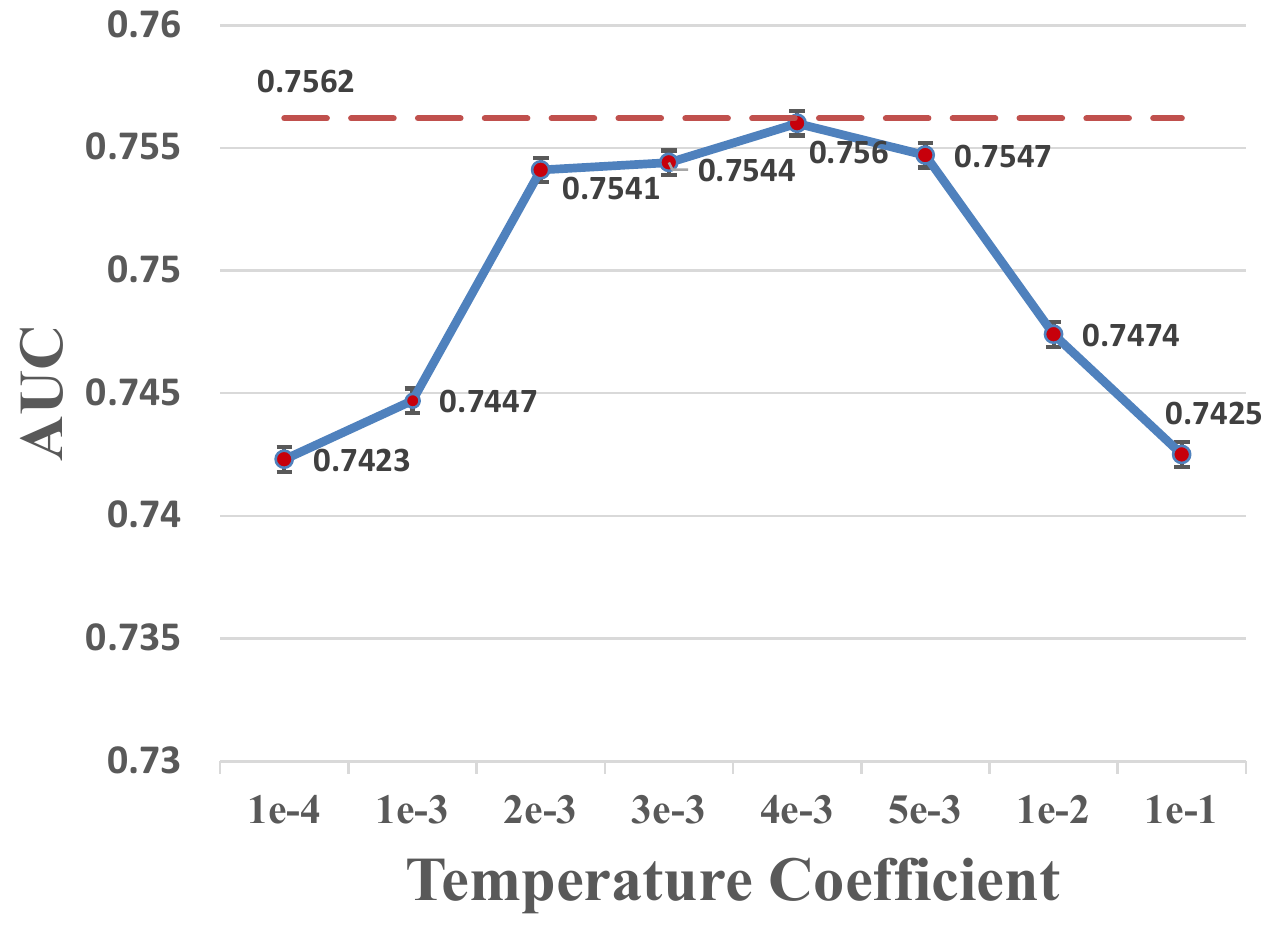}}
	\caption{\small{Impact of the temperature coefficient.}}
	\label{fig:temp_line}
\end{figure}

To demonstrate the effectiveness of temperature coefficient adaptive network mentioned in Section \ref{auto_dis} Automatic Discretization for generating feature-specific $\tau_{x_{j}}$, we compare the prediction results with global temperature coefficient $\tau$. The global temperature coefficients are searched and summarized in Figure~\ref{fig:temp_line}. We find that the optimal global temperature is around 1e-5 and 4e-3 on Criteo and AutoML datasets, respectively. However, our temperature coefficient adaptive network achieves best results (red dashed line) because it can adjust temperature coefficient adaptively for different feature based on the global field statistics feature and local input feature, obtaining more appropriate discretization distribution.

\subsubsection{Number of Meta-Embeddings}
\begin{figure}[!t]
	\centering
	\setlength{\belowcaptionskip}{-0.3cm}
	\setlength{\abovecaptionskip}{0cm}
	\subfigure[\small{Criteo}]{
		\label{fig:bucket_criteo}
		\includegraphics[width=0.21\textwidth]{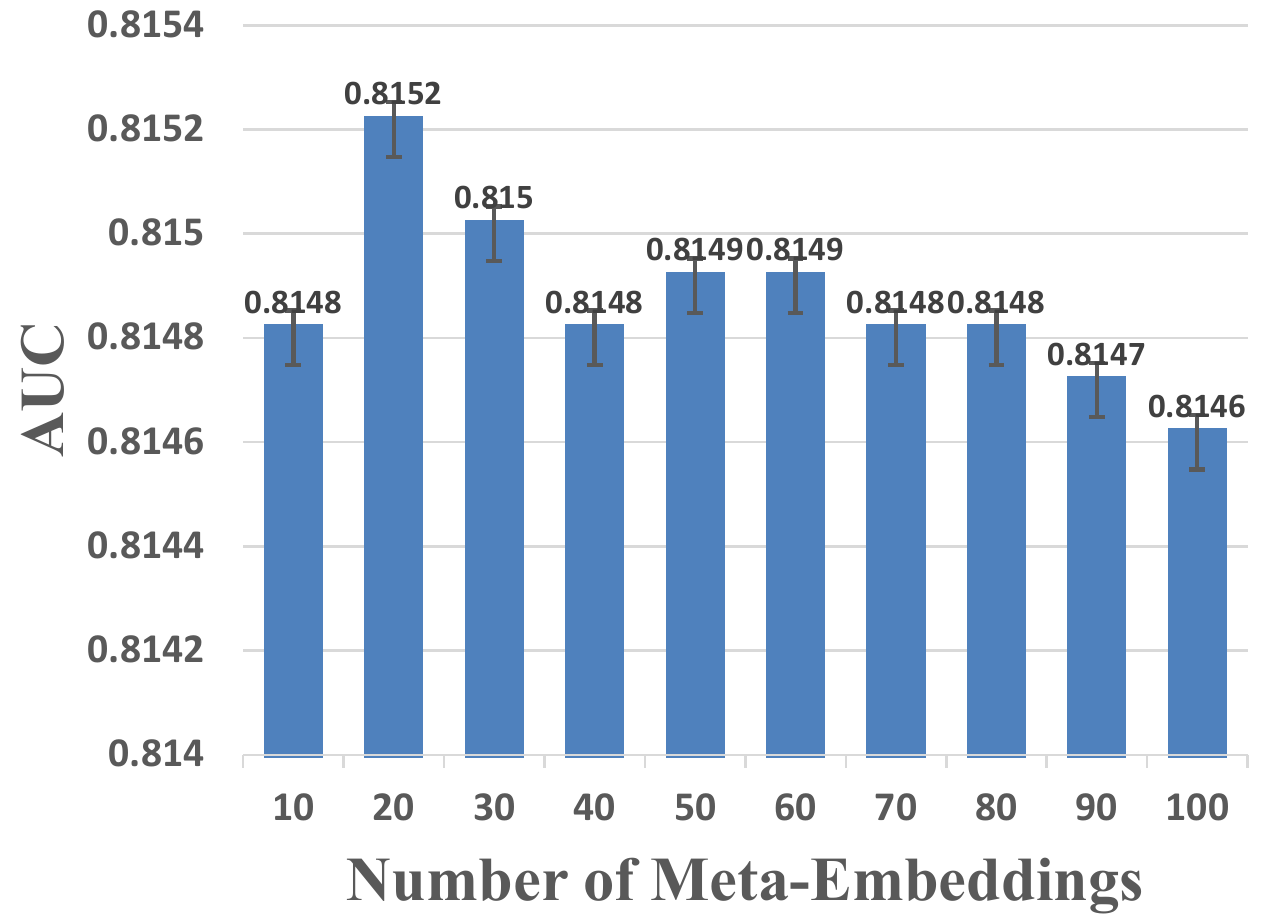}}
 	\hspace{1em}
	\subfigure[\small{AutoML}]{
		\label{fig:bucket_automl}
		\includegraphics[width=0.21\textwidth]{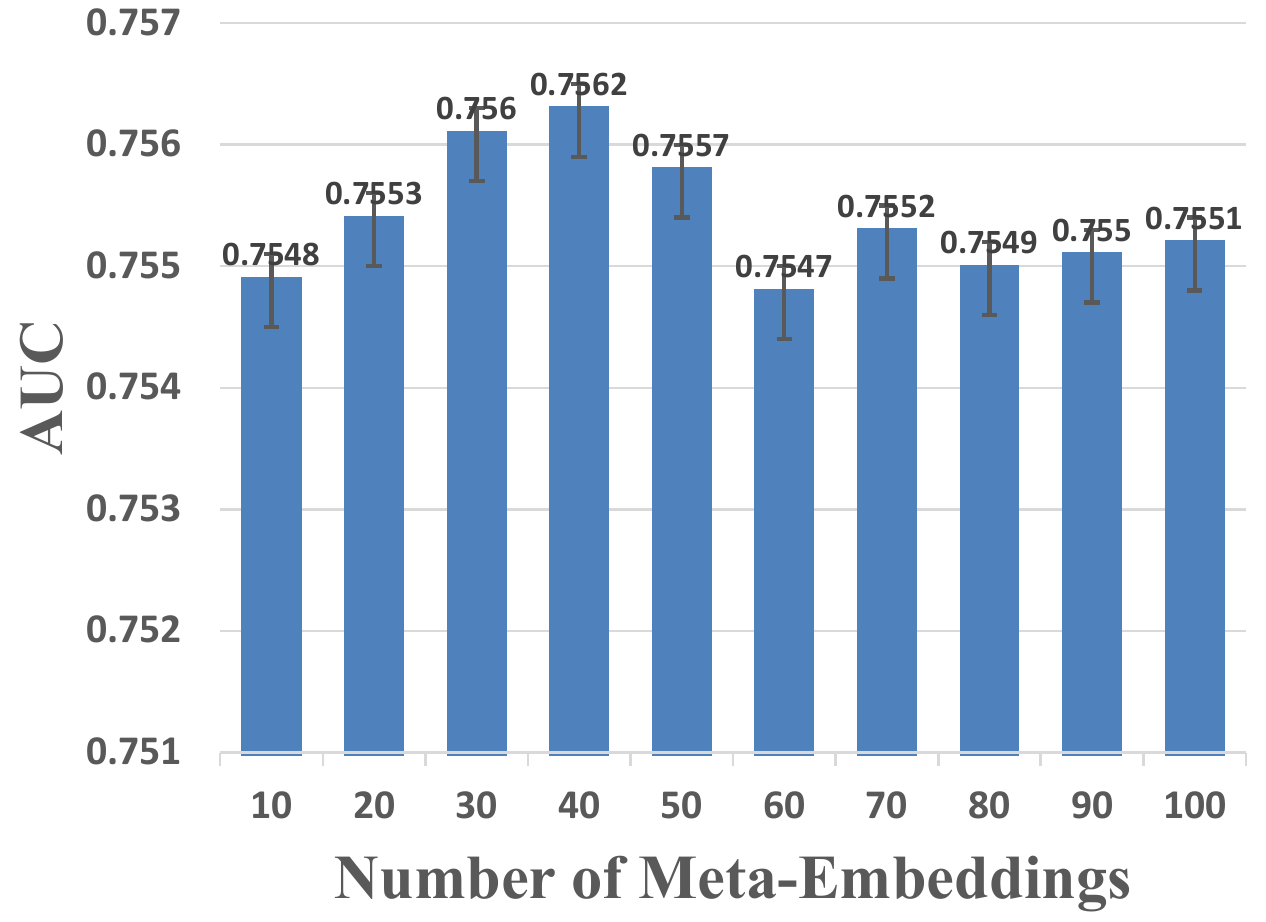}}
	\caption{\small{Impact of the number of meta-embeddings.}}
	\label{fig:bucket}
\end{figure}

We study the number of meta-embedding in this section and vary it in the range of $\{10,20,\dots,100\}$. The performance comparison is presented in Figure~\ref{fig:bucket}. It can be observed that increasing the number of meta-embeddings contributes to improving the performance substantially at the beginning. The reason lies in involving richer information in meta-embeddings. However, using too many meta-embeddings not only increases the computational burden, but also frustrates the performance. Considering both prediction effect and training efficiency, we set the number to 20 and 40 in Criteo and AutoML datasets, respectively.

\section{Related Work}
As presented in Figure~\ref{fig:Embedding_FI}, most existing deep CTR models~\cite{UBR,wide-deep,deepfm,din,increCTR}  follow an \textit{Embedding \& Feature Interaction (FI)} paradigm. Embedding module maps input features to low-dimensional embeddings. Based on such learnable embeddings, FI module models implicit/explicit low/high-order feature interactions by learning nonlinear relationship among features. Below, we will give a brief introduction of these modules in deep CTR models.

\subsection{Embedding}

As the cornerstone of the deep CTR models, Embedding module has significant impact on model performance because it takes up majority of the model parameters. Since the approaches for numerical features have been discussed in detail in Section~\ref{sec:pre}, we here give a brief introduction about research based on embedding look-up in recommendation domain. Existing researches mainly focus on designing adaptable embedding algorithms by allocating variable-length embeddings or multi-embeddings to different features, such as Mixed-dimension~\cite{mix-dimension}, NIS~\cite{NIS} and AutoEmb~\cite{autoemb}.
Another research line delves into the embedding compression~\cite{DHE} to reduce the memory footprint on the web-scale dataset.
However, these methods can only be applied in the look-up table manner for categorical features or numerical features after disretization. Few studies pay attention to the embedding learning of numerical features, which is vital in deep CTR models of the industry.

\subsection{Feature Interaction}

To capture the feature interactions effectively, different architectures are proposed in the research community of CTR prediction. 
According to different combinations of explicit and implicit feature interaction, existing CTR models can be classified into two categories: \textit{stacked structure} and \textit{parallel structure}.
Models with stacked structure first perform explicit feature interaction modeling over the embeddings and then stack a DNN to extract high-level implicit feature interactions. Representative models include FNN~\cite{Fnn}, IPNN~\cite{pnn}, PIN~\cite{pin}, FiBiNET~\cite{FiBiNet}, FGCNN~\cite{fgcnn}, AutoGroup~\cite{autogroup}, DIN~\cite{din} and DIEN~\cite{dien}.
On the contrary, models with parallel structure leverage two parallel networks to capture explicit and implicit feature interactive signals respectively, and fuse the information in the output layer. Classic members in this category are Wide \& Deep~\cite{wide-deep}, DeepFM~\cite{deepfm}, AutoFIS~\cite{autofis}, DCN~\cite{dcn}, xDeepFM~\cite{xdeepfm} and AutoInt~\cite{song2019autoint}.
In these models, implicit feature interactions are extracted via a DNN model; while for explicit feature interaction modeling, Wide \& Deep~\cite{wide-deep} uses handcrafted cross features, DeepFM and AutoFIS adopt a FM~\cite{fm} structure, DCN uses a cross network, xDeepFM employs a Compressed Interaction Network (CIN) and AutoInt leverages a multi-head self-attention network.




\section{Conclusion}
In this paper, we propose AutoDis, a pluggable embedding learning framework for numerical features in CTR prediction.
AutoDis overcomes the shortcomings of the state-of-the-art processing methods and achieves better performance.
It gains the improvement from these advantages:
(1) \textit{High model capacity}: The meta-embeddings shared by intra-field features capture global informative knowledge;
(2) \textit{End-to-end training}: The differentiable automatic discretization performs soft discretization, so that AutoDis can be optimized jointly with the ultimate goal of deep CTR models;
(3) \textit{Unique representation}: The weighted-average aggregation ensures that each feature learns a unique \textit{Continuous-But-Different} representation.
Extensive experiments on three real-world datasets and online A/B test demonstrate the effectiveness and compatibility of AutoDis.

\section*{Acknowledgement}
Weinan Zhang is supported by ``New Generation of AI 2030'' Major Project (2018AAA0100900) and National Natural Science Foundation of China (62076161, 61632017). This work is also sponsored by Huawei Innovation Research Program. We thank MindSpore~\cite{mindspore} for the partial support of this work.

\balance
\bibliographystyle{ACM-Reference-Format}
\bibliography{kdd}


\begin{thebibliography}{33}


\ifx \showCODEN    \undefined \def \showCODEN     #1{\unskip}     \fi
\ifx \showDOI      \undefined \def \showDOI       #1{#1}\fi
\ifx \showISBNx    \undefined \def \showISBNx     #1{\unskip}     \fi
\ifx \showISBNxiii \undefined \def \showISBNxiii  #1{\unskip}     \fi
\ifx \showISSN     \undefined \def \showISSN      #1{\unskip}     \fi
\ifx \showLCCN     \undefined \def \showLCCN      #1{\unskip}     \fi
\ifx \shownote     \undefined \def \shownote      #1{#1}          \fi
\ifx \showarticletitle \undefined \def \showarticletitle #1{#1}   \fi
\ifx \showURL      \undefined \def \showURL       {\relax}        \fi
\providecommand\bibfield[2]{#2}
\providecommand\bibinfo[2]{#2}
\providecommand\natexlab[1]{#1}
\providecommand\showeprint[2][]{arXiv:#2}

\bibitem[\protect\citeauthoryear{??}{min}{2020}]%
        {mindspore}
 \bibinfo{year}{2020}\natexlab{}.
\newblock \bibinfo{title}{MindSpore}.
\newblock
\newblock
\urldef\tempurl%
\url{https://www.mindspore.cn/}
\showURL{%
\tempurl}


\bibitem[\protect\citeauthoryear{Cheng, Koc, Harmsen, Shaked, Chandra, Aradhye,
  Anderson, Corrado, Chai, Ispir, Anil, Haque, Hong, Jain, Liu, and Shah}{Cheng
  et~al\mbox{.}}{2016}]%
        {wide-deep}
\bibfield{author}{\bibinfo{person}{Heng{-}Tze Cheng}, \bibinfo{person}{Levent
  Koc}, \bibinfo{person}{Jeremiah Harmsen}, \bibinfo{person}{Tal Shaked},
  \bibinfo{person}{Tushar Chandra}, \bibinfo{person}{Hrishi Aradhye},
  \bibinfo{person}{Glen Anderson}, \bibinfo{person}{Greg Corrado},
  \bibinfo{person}{Wei Chai}, \bibinfo{person}{Mustafa Ispir},
  \bibinfo{person}{Rohan Anil}, \bibinfo{person}{Zakaria Haque},
  \bibinfo{person}{Lichan Hong}, \bibinfo{person}{Vihan Jain},
  \bibinfo{person}{Xiaobing Liu}, {and} \bibinfo{person}{Hemal Shah}.}
  \bibinfo{year}{2016}\natexlab{}.
\newblock \showarticletitle{Wide {\&} Deep Learning for Recommender Systems}.
  In \bibinfo{booktitle}{\emph{Proc. Workshop Deep Learning for Recommender
  Systems}}.
\newblock


\bibitem[\protect\citeauthoryear{Covington, Adams, and Sargin}{Covington
  et~al\mbox{.}}{2016}]%
        {dnn-youtube}
\bibfield{author}{\bibinfo{person}{Paul Covington}, \bibinfo{person}{Jay
  Adams}, {and} \bibinfo{person}{Emre Sargin}.}
  \bibinfo{year}{2016}\natexlab{}.
\newblock \showarticletitle{Deep Neural Networks for YouTube Recommendations}.
  In \bibinfo{booktitle}{\emph{RecSys}}.
\newblock


\bibitem[\protect\citeauthoryear{Ginart, Naumov, Mudigere, Yang, and
  Zou}{Ginart et~al\mbox{.}}{2019}]%
        {mix-dimension}
\bibfield{author}{\bibinfo{person}{Antonio Ginart}, \bibinfo{person}{Maxim
  Naumov}, \bibinfo{person}{Dheevatsa Mudigere}, \bibinfo{person}{Jiyan Yang},
  {and} \bibinfo{person}{James Zou}.} \bibinfo{year}{2019}\natexlab{}.
\newblock \showarticletitle{Mixed dimension embeddings with application to
  memory-efficient recommendation systems}.
\newblock \bibinfo{journal}{\emph{arXiv preprint arXiv:1909.11810}}
  (\bibinfo{year}{2019}).
\newblock


\bibitem[\protect\citeauthoryear{Grabczewski and Jankowski}{Grabczewski and
  Jankowski}{2005}]%
        {DT-FS}
\bibfield{author}{\bibinfo{person}{Krzysztof Grabczewski} {and}
  \bibinfo{person}{Norbert Jankowski}.} \bibinfo{year}{2005}\natexlab{}.
\newblock \showarticletitle{Feature Selection with Decision Tree Criterion}. In
  \bibinfo{booktitle}{\emph{HIS}}. \bibinfo{publisher}{{IEEE} Computer
  Society}, \bibinfo{pages}{212--217}.
\newblock


\bibitem[\protect\citeauthoryear{Gu, Ding, Wang, Zou, Liu, and Yin}{Gu
  et~al\mbox{.}}{2020}]%
        {DMT}
\bibfield{author}{\bibinfo{person}{Yulong Gu}, \bibinfo{person}{Zhuoye Ding},
  \bibinfo{person}{Shuaiqiang Wang}, \bibinfo{person}{Lixin Zou},
  \bibinfo{person}{Yiding Liu}, {and} \bibinfo{person}{Dawei Yin}.}
  \bibinfo{year}{2020}\natexlab{}.
\newblock \showarticletitle{Deep Multifaceted Transformers for Multi-objective
  Ranking in Large-Scale E-commerce Recommender Systems}. In
  \bibinfo{booktitle}{\emph{CIKM}}. \bibinfo{publisher}{{ACM}},
  \bibinfo{pages}{2493--2500}.
\newblock


\bibitem[\protect\citeauthoryear{Guo, Tang, Ye, Li, and He}{Guo
  et~al\mbox{.}}{2017}]%
        {deepfm}
\bibfield{author}{\bibinfo{person}{Huifeng Guo}, \bibinfo{person}{Ruiming
  Tang}, \bibinfo{person}{Yunming Ye}, \bibinfo{person}{Zhenguo Li}, {and}
  \bibinfo{person}{Xiuqiang He}.} \bibinfo{year}{2017}\natexlab{}.
\newblock \showarticletitle{DeepFM: {A} Factorization-Machine based Neural
  Network for {CTR} Prediction}. In \bibinfo{booktitle}{\emph{IJCAI}}.
\newblock


\bibitem[\protect\citeauthoryear{He, Zhang, Ren, and Sun}{He
  et~al\mbox{.}}{2015}]%
        {he2015delving}
\bibfield{author}{\bibinfo{person}{Kaiming He}, \bibinfo{person}{Xiangyu
  Zhang}, \bibinfo{person}{Shaoqing Ren}, {and} \bibinfo{person}{Jian Sun}.}
  \bibinfo{year}{2015}\natexlab{}.
\newblock \showarticletitle{Delving deep into rectifiers: Surpassing
  human-level performance on imagenet classification}. In
  \bibinfo{booktitle}{\emph{ICCV}}. \bibinfo{pages}{1026--1034}.
\newblock


\bibitem[\protect\citeauthoryear{He, Pan, Jin, Xu, Liu, Xu, Shi, Atallah,
  Herbrich, Bowers, and Candela}{He et~al\mbox{.}}{2014}]%
        {fb_practical}
\bibfield{author}{\bibinfo{person}{Xinran He}, \bibinfo{person}{Junfeng Pan},
  \bibinfo{person}{Ou Jin}, \bibinfo{person}{Tianbing Xu}, \bibinfo{person}{Bo
  Liu}, \bibinfo{person}{Tao Xu}, \bibinfo{person}{Yanxin Shi},
  \bibinfo{person}{Antoine Atallah}, \bibinfo{person}{Ralf Herbrich},
  \bibinfo{person}{Stuart Bowers}, {and}
  \bibinfo{person}{Joaquin~Qui{\~{n}}onero Candela}.}
  \bibinfo{year}{2014}\natexlab{}.
\newblock \showarticletitle{Practical Lessons from Predicting Clicks on Ads at
  Facebook}. In \bibinfo{booktitle}{\emph{ADKDD}}. \bibinfo{publisher}{{ACM}}.
\newblock


\bibitem[\protect\citeauthoryear{Huang, Zhang, and Zhang}{Huang
  et~al\mbox{.}}{2019}]%
        {FiBiNet}
\bibfield{author}{\bibinfo{person}{Tongwen Huang}, \bibinfo{person}{Zhiqi
  Zhang}, {and} \bibinfo{person}{Junlin Zhang}.}
  \bibinfo{year}{2019}\natexlab{}.
\newblock \showarticletitle{FiBiNET: combining feature importance and bilinear
  feature interaction for click-through rate prediction}. In
  \bibinfo{booktitle}{\emph{RecSys}}. \bibinfo{publisher}{{ACM}},
  \bibinfo{pages}{169--177}.
\newblock


\bibitem[\protect\citeauthoryear{Joglekar, Li, Adams, Khaitan, and Le}{Joglekar
  et~al\mbox{.}}{2019}]%
        {NIS}
\bibfield{author}{\bibinfo{person}{Manas~R. Joglekar}, \bibinfo{person}{Cong
  Li}, \bibinfo{person}{Jay~K. Adams}, \bibinfo{person}{Pranav Khaitan}, {and}
  \bibinfo{person}{Quoc~V. Le}.} \bibinfo{year}{2019}\natexlab{}.
\newblock \showarticletitle{Neural Input Search for Large Scale Recommendation
  Models}.
\newblock \bibinfo{journal}{\emph{CoRR}}  \bibinfo{volume}{abs/1907.04471}
  (\bibinfo{year}{2019}).
\newblock
\showeprint[arxiv]{1907.04471}
\urldef\tempurl%
\url{http://arxiv.org/abs/1907.04471}
\showURL{%
\tempurl}


\bibitem[\protect\citeauthoryear{Juan, Zhuang, Chin, and Lin}{Juan
  et~al\mbox{.}}{2016}]%
        {ffm}
\bibfield{author}{\bibinfo{person}{Yu{-}Chin Juan}, \bibinfo{person}{Yong
  Zhuang}, \bibinfo{person}{Wei{-}Sheng Chin}, {and}
  \bibinfo{person}{Chih{-}Jen Lin}.} \bibinfo{year}{2016}\natexlab{}.
\newblock \showarticletitle{Field-aware Factorization Machines for {CTR}
  Prediction}. In \bibinfo{booktitle}{\emph{RecSys}}.
\newblock


\bibitem[\protect\citeauthoryear{Kang, Cheng, Yao, Yi, Chen, Hong, and
  Chi}{Kang et~al\mbox{.}}{2020}]%
        {DHE}
\bibfield{author}{\bibinfo{person}{Wang{-}Cheng Kang},
  \bibinfo{person}{Derek~Zhiyuan Cheng}, \bibinfo{person}{Tiansheng Yao},
  \bibinfo{person}{Xinyang Yi}, \bibinfo{person}{Ting Chen},
  \bibinfo{person}{Lichan Hong}, {and} \bibinfo{person}{Ed~H. Chi}.}
  \bibinfo{year}{2020}\natexlab{}.
\newblock \showarticletitle{Deep Hash Embedding for Large-Vocab Categorical
  Feature Representations}.
\newblock \bibinfo{journal}{\emph{CoRR}}  \bibinfo{volume}{abs/2010.10784}
  (\bibinfo{year}{2020}).
\newblock


\bibitem[\protect\citeauthoryear{Ke, Xu, Zhang, Bian, and Liu}{Ke
  et~al\mbox{.}}{2019}]%
        {DeepGBM}
\bibfield{author}{\bibinfo{person}{Guolin Ke}, \bibinfo{person}{Zhenhui Xu},
  \bibinfo{person}{Jia Zhang}, \bibinfo{person}{Jiang Bian}, {and}
  \bibinfo{person}{Tie{-}Yan Liu}.} \bibinfo{year}{2019}\natexlab{}.
\newblock \showarticletitle{DeepGBM: {A} Deep Learning Framework Distilled by
  {GBDT} for Online Prediction Tasks}. In \bibinfo{booktitle}{\emph{SIGKDD}}.
  \bibinfo{publisher}{{ACM}}, \bibinfo{pages}{384--394}.
\newblock


\bibitem[\protect\citeauthoryear{Kingma and Ba}{Kingma and Ba}{2014}]%
        {adam}
\bibfield{author}{\bibinfo{person}{Diederik~P Kingma} {and}
  \bibinfo{person}{Jimmy Ba}.} \bibinfo{year}{2014}\natexlab{}.
\newblock \showarticletitle{Adam: A method for stochastic optimization}.
\newblock \bibinfo{journal}{\emph{arXiv preprint arXiv:1412.6980}}
  (\bibinfo{year}{2014}).
\newblock


\bibitem[\protect\citeauthoryear{Lian, Zhou, Zhang, Chen, Xie, and Sun}{Lian
  et~al\mbox{.}}{2018}]%
        {xdeepfm}
\bibfield{author}{\bibinfo{person}{Jianxun Lian}, \bibinfo{person}{Xiaohuan
  Zhou}, \bibinfo{person}{Fuzheng Zhang}, \bibinfo{person}{Zhongxia Chen},
  \bibinfo{person}{Xing Xie}, {and} \bibinfo{person}{Guangzhong Sun}.}
  \bibinfo{year}{2018}\natexlab{}.
\newblock \showarticletitle{{xDeepFM}: Combining Explicit and Implicit Feature
  Interactions for Recommender Systems}. In \bibinfo{booktitle}{\emph{SIGKDD}}.
\newblock


\bibitem[\protect\citeauthoryear{Liu, Tang, Chen, Yu, Guo, and Zhang}{Liu
  et~al\mbox{.}}{2019}]%
        {fgcnn}
\bibfield{author}{\bibinfo{person}{Bin Liu}, \bibinfo{person}{Ruiming Tang},
  \bibinfo{person}{Yingzhi Chen}, \bibinfo{person}{Jinkai Yu},
  \bibinfo{person}{Huifeng Guo}, {and} \bibinfo{person}{Yuzhou Zhang}.}
  \bibinfo{year}{2019}\natexlab{}.
\newblock \showarticletitle{Feature Generation by Convolutional Neural Network
  for Click-Through Rate Prediction}. In \bibinfo{booktitle}{\emph{WWW}}.
\newblock


\bibitem[\protect\citeauthoryear{Liu, Xue, Guo, Tang, Zafeiriou, He, and
  Li}{Liu et~al\mbox{.}}{2020a}]%
        {autogroup}
\bibfield{author}{\bibinfo{person}{Bin Liu}, \bibinfo{person}{Niannan Xue},
  \bibinfo{person}{Huifeng Guo}, \bibinfo{person}{Ruiming Tang},
  \bibinfo{person}{Stefanos Zafeiriou}, \bibinfo{person}{Xiuqiang He}, {and}
  \bibinfo{person}{Zhenguo Li}.} \bibinfo{year}{2020}\natexlab{a}.
\newblock \showarticletitle{AutoGroup: Automatic Feature Grouping for Modelling
  Explicit High-Order Feature Interactions in {CTR} Prediction}. In
  \bibinfo{booktitle}{\emph{SIGIR}}. \bibinfo{publisher}{{ACM}}.
\newblock


\bibitem[\protect\citeauthoryear{Liu, Zhu, Li, Zhang, Lai, Tang, He, Li, and
  Yu}{Liu et~al\mbox{.}}{2020b}]%
        {autofis}
\bibfield{author}{\bibinfo{person}{Bin Liu}, \bibinfo{person}{Chenxu Zhu},
  \bibinfo{person}{Guilin Li}, \bibinfo{person}{Weinan Zhang},
  \bibinfo{person}{Jincai Lai}, \bibinfo{person}{Ruiming Tang},
  \bibinfo{person}{Xiuqiang He}, \bibinfo{person}{Zhenguo Li}, {and}
  \bibinfo{person}{Yong Yu}.} \bibinfo{year}{2020}\natexlab{b}.
\newblock \showarticletitle{AutoFIS: Automatic Feature Interaction Selection in
  Factorization Models for Click-Through Rate Prediction}. In
  \bibinfo{booktitle}{\emph{SIGKDD}}.
\newblock


\bibitem[\protect\citeauthoryear{McMahan, Holt, Sculley, Young, Ebner, Grady,
  Nie, Phillips, Davydov, Golovin, et~al\mbox{.}}{McMahan
  et~al\mbox{.}}{2013}]%
        {ftrl}
\bibfield{author}{\bibinfo{person}{H~Brendan McMahan}, \bibinfo{person}{Gary
  Holt}, \bibinfo{person}{David Sculley}, \bibinfo{person}{Michael Young},
  \bibinfo{person}{Dietmar Ebner}, \bibinfo{person}{Julian Grady},
  \bibinfo{person}{Lan Nie}, \bibinfo{person}{Todd Phillips},
  \bibinfo{person}{Eugene Davydov}, \bibinfo{person}{Daniel Golovin},
  {et~al\mbox{.}}} \bibinfo{year}{2013}\natexlab{}.
\newblock \showarticletitle{Ad click prediction: a view from the trenches}. In
  \bibinfo{booktitle}{\emph{SIGKDD}}. ACM, \bibinfo{pages}{1222--1230}.
\newblock


\bibitem[\protect\citeauthoryear{Naumov, Mudigere, Shi, Huang, Sundaraman,
  Park, Wang, Gupta, Wu, Azzolini, Dzhulgakov, Mallevich, Cherniavskii, Lu,
  Krishnamoorthi, Yu, Kondratenko, Pereira, Chen, Chen, Rao, Jia, Xiong, and
  Smelyanskiy}{Naumov et~al\mbox{.}}{2019}]%
        {DLRM}
\bibfield{author}{\bibinfo{person}{Maxim Naumov}, \bibinfo{person}{Dheevatsa
  Mudigere}, \bibinfo{person}{Hao{-}Jun~Michael Shi}, \bibinfo{person}{Jianyu
  Huang}, \bibinfo{person}{Narayanan Sundaraman}, \bibinfo{person}{Jongsoo
  Park}, \bibinfo{person}{Xiaodong Wang}, \bibinfo{person}{Udit Gupta},
  \bibinfo{person}{Carole{-}Jean Wu}, \bibinfo{person}{Alisson~G. Azzolini},
  \bibinfo{person}{Dmytro Dzhulgakov}, \bibinfo{person}{Andrey Mallevich},
  \bibinfo{person}{Ilia Cherniavskii}, \bibinfo{person}{Yinghai Lu},
  \bibinfo{person}{Raghuraman Krishnamoorthi}, \bibinfo{person}{Ansha Yu},
  \bibinfo{person}{Volodymyr Kondratenko}, \bibinfo{person}{Stephanie Pereira},
  \bibinfo{person}{Xianjie Chen}, \bibinfo{person}{Wenlin Chen},
  \bibinfo{person}{Vijay Rao}, \bibinfo{person}{Bill Jia},
  \bibinfo{person}{Liang Xiong}, {and} \bibinfo{person}{Misha Smelyanskiy}.}
  \bibinfo{year}{2019}\natexlab{}.
\newblock \showarticletitle{Deep Learning Recommendation Model for
  Personalization and Recommendation Systems}.
\newblock \bibinfo{journal}{\emph{CoRR}}  \bibinfo{volume}{abs/1906.00091}
  (\bibinfo{year}{2019}).
\newblock


\bibitem[\protect\citeauthoryear{Qin, Zhang, Wu, Jin, Fang, and Yu}{Qin
  et~al\mbox{.}}{2020}]%
        {UBR}
\bibfield{author}{\bibinfo{person}{Jiarui Qin}, \bibinfo{person}{Weinan Zhang},
  \bibinfo{person}{Xin Wu}, \bibinfo{person}{Jiarui Jin},
  \bibinfo{person}{Yuchen Fang}, {and} \bibinfo{person}{Yong Yu}.}
  \bibinfo{year}{2020}\natexlab{}.
\newblock \showarticletitle{User Behavior Retrieval for Click-Through Rate
  Prediction}. In \bibinfo{booktitle}{\emph{SIGIR}}.
  \bibinfo{publisher}{{ACM}}.
\newblock


\bibitem[\protect\citeauthoryear{Qu, Cai, Ren, Zhang, Yu, Wen, and Wang}{Qu
  et~al\mbox{.}}{2016}]%
        {pnn}
\bibfield{author}{\bibinfo{person}{Yanru Qu}, \bibinfo{person}{Han Cai},
  \bibinfo{person}{Kan Ren}, \bibinfo{person}{Weinan Zhang},
  \bibinfo{person}{Yong Yu}, \bibinfo{person}{Ying Wen}, {and}
  \bibinfo{person}{Jun Wang}.} \bibinfo{year}{2016}\natexlab{}.
\newblock \showarticletitle{Product-based neural networks for user response
  prediction}. In \bibinfo{booktitle}{\emph{ICDM}}. IEEE,
  \bibinfo{pages}{1149--1154}.
\newblock


\bibitem[\protect\citeauthoryear{Qu, Fang, Zhang, Tang, Niu, Guo, Yu, and
  He}{Qu et~al\mbox{.}}{2019}]%
        {pin}
\bibfield{author}{\bibinfo{person}{Yanru Qu}, \bibinfo{person}{Bohui Fang},
  \bibinfo{person}{Weinan Zhang}, \bibinfo{person}{Ruiming Tang},
  \bibinfo{person}{Minzhe Niu}, \bibinfo{person}{Huifeng Guo},
  \bibinfo{person}{Yong Yu}, {and} \bibinfo{person}{Xiuqiang He}.}
  \bibinfo{year}{2019}\natexlab{}.
\newblock \showarticletitle{Product-based Neural Networks for User Response
  Prediction over Multi-field Categorical Data}.
\newblock \bibinfo{journal}{\emph{{ACM} Trans. Inf. Syst.}}
  \bibinfo{volume}{37}, \bibinfo{number}{1} (\bibinfo{year}{2019}),
  \bibinfo{pages}{5:1--5:35}.
\newblock


\bibitem[\protect\citeauthoryear{Rendle}{Rendle}{2010}]%
        {fm}
\bibfield{author}{\bibinfo{person}{Steffen Rendle}.}
  \bibinfo{year}{2010}\natexlab{}.
\newblock \showarticletitle{Factorization Machines}. In
  \bibinfo{booktitle}{\emph{Proc. IEEE Int. Conf. Data Mining}}.
\newblock


\bibitem[\protect\citeauthoryear{Song, Shi, Xiao, Duan, Xu, Zhang, and
  Tang}{Song et~al\mbox{.}}{2019}]%
        {song2019autoint}
\bibfield{author}{\bibinfo{person}{Weiping Song}, \bibinfo{person}{Chence Shi},
  \bibinfo{person}{Zhiping Xiao}, \bibinfo{person}{Zhijian Duan},
  \bibinfo{person}{Yewen Xu}, \bibinfo{person}{Ming Zhang}, {and}
  \bibinfo{person}{Jian Tang}.} \bibinfo{year}{2019}\natexlab{}.
\newblock \showarticletitle{Autoint: Automatic feature interaction learning via
  self-attentive neural networks}. In \bibinfo{booktitle}{\emph{CIKM}}.
  \bibinfo{pages}{1161--1170}.
\newblock


\bibitem[\protect\citeauthoryear{Van~der Maaten and Hinton}{Van~der Maaten and
  Hinton}{2008}]%
        {van2008visualizing}
\bibfield{author}{\bibinfo{person}{Laurens Van~der Maaten} {and}
  \bibinfo{person}{Geoffrey Hinton}.} \bibinfo{year}{2008}\natexlab{}.
\newblock \showarticletitle{Visualizing data using t-SNE.}
\newblock \bibinfo{journal}{\emph{Journal of machine learning research}}
  \bibinfo{volume}{9}, \bibinfo{number}{11} (\bibinfo{year}{2008}).
\newblock


\bibitem[\protect\citeauthoryear{Wang, Fu, Fu, and Wang}{Wang
  et~al\mbox{.}}{2017}]%
        {dcn}
\bibfield{author}{\bibinfo{person}{Ruoxi Wang}, \bibinfo{person}{Bin Fu},
  \bibinfo{person}{Gang Fu}, {and} \bibinfo{person}{Mingliang Wang}.}
  \bibinfo{year}{2017}\natexlab{}.
\newblock \showarticletitle{Deep {\&} Cross Network for Ad Click Predictions}.
  In \bibinfo{booktitle}{\emph{ADKDD}}. \bibinfo{publisher}{{ACM}},
  \bibinfo{pages}{12:1--12:7}.
\newblock


\bibitem[\protect\citeauthoryear{Wang, Guo, Tang, Liu, and He}{Wang
  et~al\mbox{.}}{2020}]%
        {increCTR}
\bibfield{author}{\bibinfo{person}{Yichao Wang}, \bibinfo{person}{Huifeng Guo},
  \bibinfo{person}{Ruiming Tang}, \bibinfo{person}{Zhirong Liu}, {and}
  \bibinfo{person}{Xiuqiang He}.} \bibinfo{year}{2020}\natexlab{}.
\newblock \showarticletitle{A Practical Incremental Method to Train Deep {CTR}
  Models}.
\newblock \bibinfo{journal}{\emph{CoRR}}  \bibinfo{volume}{abs/2009.02147}
  (\bibinfo{year}{2020}).
\newblock


\bibitem[\protect\citeauthoryear{Zhang, Du, and Wang}{Zhang
  et~al\mbox{.}}{2016}]%
        {Fnn}
\bibfield{author}{\bibinfo{person}{Weinan Zhang}, \bibinfo{person}{Tianming
  Du}, {and} \bibinfo{person}{Jun Wang}.} \bibinfo{year}{2016}\natexlab{}.
\newblock \showarticletitle{Deep learning over multi-field categorical data}.
  In \bibinfo{booktitle}{\emph{ECIR}}. Springer, \bibinfo{pages}{45--57}.
\newblock


\bibitem[\protect\citeauthoryear{Zhao, Wang, Chen, Zheng, Liu, and Tang}{Zhao
  et~al\mbox{.}}{2020}]%
        {autoemb}
\bibfield{author}{\bibinfo{person}{Xiangyu Zhao}, \bibinfo{person}{Chong Wang},
  \bibinfo{person}{Ming Chen}, \bibinfo{person}{Xudong Zheng},
  \bibinfo{person}{Xiaobing Liu}, {and} \bibinfo{person}{Jiliang Tang}.}
  \bibinfo{year}{2020}\natexlab{}.
\newblock \showarticletitle{AutoEmb: Automated Embedding Dimensionality Search
  in Streaming Recommendations}.
\newblock \bibinfo{journal}{\emph{arXiv preprint arXiv:2002.11252}}
  (\bibinfo{year}{2020}).
\newblock


\bibitem[\protect\citeauthoryear{Zhou, Mou, Fan, Pi, Bian, Zhou, Zhu, and
  Gai}{Zhou et~al\mbox{.}}{2019}]%
        {dien}
\bibfield{author}{\bibinfo{person}{Guorui Zhou}, \bibinfo{person}{Na Mou},
  \bibinfo{person}{Ying Fan}, \bibinfo{person}{Qi Pi}, \bibinfo{person}{Weijie
  Bian}, \bibinfo{person}{Chang Zhou}, \bibinfo{person}{Xiaoqiang Zhu}, {and}
  \bibinfo{person}{Kun Gai}.} \bibinfo{year}{2019}\natexlab{}.
\newblock \showarticletitle{Deep Interest Evolution Network for Click-Through
  Rate Prediction}. In \bibinfo{booktitle}{\emph{AAAI}}.
  \bibinfo{pages}{5941--5948}.
\newblock


\bibitem[\protect\citeauthoryear{Zhou, Zhu, Song, Fan, Zhu, Ma, Yan, Jin, Li,
  and Gai}{Zhou et~al\mbox{.}}{2018}]%
        {din}
\bibfield{author}{\bibinfo{person}{Guorui Zhou}, \bibinfo{person}{Xiaoqiang
  Zhu}, \bibinfo{person}{Chengru Song}, \bibinfo{person}{Ying Fan},
  \bibinfo{person}{Han Zhu}, \bibinfo{person}{Xiao Ma},
  \bibinfo{person}{Yanghui Yan}, \bibinfo{person}{Junqi Jin},
  \bibinfo{person}{Han Li}, {and} \bibinfo{person}{Kun Gai}.}
  \bibinfo{year}{2018}\natexlab{}.
\newblock \showarticletitle{Deep Interest Network for Click-Through Rate
  Prediction}. In \bibinfo{booktitle}{\emph{SIGKDD}}.
  \bibinfo{pages}{1059--1068}.
\newblock


\end{thebibliography}

\end{document}